%% file: main-arxiv.tex
\documentclass[conference]{IEEEtran}
\IEEEoverridecommandlockouts
\usepackage{float}

\usepackage{tikz}
\usepackage{amsmath}
\usepackage{caption}
\usepackage{algorithm}
\usepackage{algorithmic}
\usepackage{placeins}
\usepackage{float}
\usepackage{amssymb}
\usepackage{filecontents}

\makeatletter
\renewcommand{\footnoterule}{%
  \kern -3pt
  \hrule width \columnwidth
  \kern 2.6pt
}
\makeatother

\begin{document}
%-------------------------------------------------------------------------------

%don't want date printed
\date{}

% make title bold and 14 pt font (Latex default is non-bold, 16 pt)
\title{Secure Transformer Generation with Encrypted KV-Cache Reuse}

%for single author (just remove % characters)
% \author{
% {\rm Hedong Zhang}\\
% University of Central Florida
% \and
% {\rm Neusha Javidnia}\\
% University of California, San Diego
% \and
% {\rm Shweta Pardeshi}\\
% University of California, San Diego
% \and
% {\rm Qian Lou\thanks{Corresponding author: qian.lou@ucf.edu.}}\\
% University of Central Florida
% \and
% {\rm Farinaz Koushanfar}\\
% University of California, San Diego
% } % end author

\author{
\IEEEauthorblockN{
Hedong Zhang$^{1}$,
Neusha Javidnia$^{2}$,
Shweta Pardeshi$^{2}$,
Qian Lou$^{1}$,
Farinaz Koushanfar$^{2}$
}
\thanks{
$^{1}$ University of Central Florida.
$^{2}$ University of California, San Diego.
Corresponding author: Qian Lou (qian.lou@ucf.edu).
}
}

\maketitle
\thispagestyle{plain}
%-------------------------------------------------------------------------------
\begin{abstract}
%-------------------------------------------------------------------------------
The widespread deployment of cloud-hosted generative models raises a fundamental challenge: enabling efficient autoregressive generation while preserving the privacy of both user prompts and model parameters in untrusted environments. We address this challenge in a client–server setting where an untrusted server hosts an autoregressive Transformer and the client requires cryptographic protection for both inputs and inference. We present \textit{CryptoGen}, the first system to enable scalable privacy-preserving neural generation with persistent encrypted key–value (KV) cache reuse. Discriminative-task secure inference systems incur quadratic latency and memory growth when adapted to autoregressive decoding due to the lack of native encrypted KV-cache support. In contrast, \textit{CryptoGen} achieves near-linear scaling by securely reusing and updating encrypted KV caches throughout generation. \textit{CryptoGen} integrates homomorphic encryption and secret sharing to support both prefilling and generation. Key techniques include a unified encrypted KV-cache framework, heterogeneous SIMD encodings for different phases, optimized cipher–cipher matrix–matrix and matrix–vector operations, and efficient noise refresh and ciphertext concatenation mechanisms. Evaluation on generative Transformer models trained on WikiText-2, PTB, and LAMBADA shows that for input lengths of 128–512 tokens, \textit{CryptoGen} achieves 4.4×–7.6× lower per-token latency than state-of-the-art discriminative secure inference systems, while maintaining near-linear latency and memory scaling, with advantages increasing for longer sequences. \textit{CryptoGen} is released as an open-source library.
\end{abstract}

%-------------------------------------------------------------------------------
\section{Introduction}
%-------------------------------------------------------------------------------
\input{contents_usenix/1_intro}
\section{Background}
\input{contents_usenix/2_background}
\section*{Threat Model}
\input{contents_usenix/threat_model}
%-------------------------------------------------------------------------------
\section{Motivation}
\input{contents_usenix/3_motivation}
\section{CryptoGen}

\input{contents_usenix/5_method}

%-------------------------------------------------------------------------------
\section{Experimental Methodologies}
%-------------------------------------------------------------------------------
\input{contents_usenix/Experimental_Methodologies}

%-------------------------------------------------------------------------------
\section{Results}
%-------------------------------------------------------------------------------
\input{contents_usenix/6_result}

%-------------------------------------------------------------------------------
\section{Conclusion}
%-------------------------------------------------------------------------------
\input{contents_usenix/7_conclusion}

%-------------------------------------------------------------------------------

% optional clearing of the page
%\cleardoublepage
% \section*{Ethical Considerations}
% \noindent\textbf{Ethical Considerations.} This work studies the systems and cryptographic aspects of privacy-preserving autoregressive Transformer inference. It does not modify model training objectives, datasets, or generation policies. \\
% \noindent\textbf{User Privacy.} CryptoGen prevents the untrusted server from observing user prompts, intermediate states, or past tokens by executing both prefilling and autoregressive decoding over encrypted data. The encrypted KV-cache reuse mechanism avoids repeated reprocessing or plaintext exposure of historical context, reducing inference-time leakage risks.\\
% \noindent\textbf{Threat Model. } CryptoGen assumes a standard semi-honest adversarial model. It does not protect against malicious adversaries that arbitrarily deviate from the protocol. Extending the system to stronger threat models (e.g., malicious security) is left to future work. 
% optional clearing of the page
%\clearpage

% \section*{Open Science}
% https://anonymous.4open.science/r/CrypGen-2FE6/README.md
% optional clearing of the page
%\clearpage
% \bibliographystyle{plainurl}
\bibliographystyle{IEEEtran}
\bibliography{main,lou}
\clearpage
\appendix
\section{Appendix}
\input{contents_usenix/appendix}
%%%%%%%%%%%%%%%%%%%%%%%%%%%%%%%%%%%%%%%%%%%%%%%%%%%%%%%%%%%%%%%%%%%%%%%%%%%%%%%%
\end{document}

%% file: contents_usenix/1_intro.tex
Fueled by the massive influx of data, sophisticated algorithms, and extensive computational resources, modern generative machine learning has achieved remarkable success across various domains, including medical diagnosis and explanation, electronic health record analysis, and credit risk assessment\cite{Abdullah2021ARO,kozodoi2022fairness,rajkomar2018scalable,wang2025survey}. These models are typically realized through autoregressive Transformers, a class of neural network architectures that leverage attention mechanisms to capture long-range dependencies and contextual relationships among input tokens\cite{vaswani2017attention}. In a generative setting, computation proceeds in two distinct yet interdependent phases\cite{radford2019language}. The \textit{prefilling} phase encodes the user and system prompts to establish contextual embeddings, while the \textit{generation} phase produces output tokens autoregressively—each new token conditioned on both the prefilled context and all previously generated tokens.

A central question arises when deploying these powerful models: how can one enable secure access to generative AI services without compromising user or model privacy? One approach is to distribute the model for local execution on client devices\cite{alizadeh2024llm}\cite{touvron2023llama}. However, this option is undesirable for at least two reasons. First, once released, the model can be copied freely, eliminating the model owner’s opportunity to monetize or control its usage. Second, such models may have been trained on sensitive or proprietary data, and distributing them risks exposing private information about data owners—potentially violating regulations such as HIPAA in healthcare contexts\cite{moore2019review, fredrikson2015model,hu2022membership}.
Alternatively, providers may offer the model as a cloud-based web service, allowing clients to send prompts and receive generated outputs over the network\cite{he2024emerged}. Yet this approach also raises serious privacy and liability concerns. Users must trust the service not to inspect, store, or misuse their prompts, while data custodians may themselves be reluctant to process user inputs on their servers to avoid legal liability in the event of data breaches. Neither option offers a satisfactory privacy solution for both parties.

Our work aims to resolve this conundrum of \textit{secure neural network generation}. More concretely, we seek to enable a protocol in which the user obtains the model’s generation results without revealing their private inputs, while the service provider gains no knowledge of these inputs and simultaneously preserves the confidentiality of its model parameters\cite{du2001secure}. Such a design provides dual-sided privacy guarantees—protecting both the user’s data and the model owner’s intellectual property. 

To achieve this vision, we leverage powerful tools from modern cryptography, particularly Fully Homomorphic Encryption (FHE) and Multi-Party Computation (MPC). FHE enables computation to be performed directly on encrypted data without decryption, allowing clients to delegate heavy computation to untrusted servers while preserving data confidentiality\cite{gentry2009fully,chen2017simple,fan2012somewhat}. In contrast, MPC distributes the client’s input among multiple parties—often via secret sharing—and executes the computation collaboratively through secure protocols\cite{yao1982protocols}. Prior work has shown that hybrid HE--MPC designs are particularly effective for \emph{discriminative} secure inference: FHE efficiently supports linear layers, while MPC evaluates non-linear functions (e.g., activations and Softmax) more naturally, outperforming using either technique alone\cite{juvekar2018gazelle,hao2022iron,pang2024bolt,moon2024thor,lu2023bumblebee}.

However, secure \emph{generation} is qualitatively harder than secure classification. Existing state-of-the-art systems---including Gazelle\cite{juvekar2018gazelle}, IRON\cite{hao2022iron}, BOLT\cite{pang2024bolt}, and THOR\cite{moon2024thor}---are optimized for discriminative (encoder-style) inference, where the full input is known and computation can be executed in a single batched pass that densely fills SIMD slots. %Some of these works evaluate generative models, but typically only for discriminative objectives or for producing a \emph{single} next-token output, which remains closer to prefilling than to token-by-token decoding.
In contrast, auto-regressive decoding is stateful and sequential: each new token depends on an ever-growing Key--Value (KV) cache from prior steps\cite{radford2019language}. This creates two fundamental obstacles when directly applying discriminative-style encrypted inference to generation.
(1) \textbf{No encrypted KV-cache reuse:} without an efficient encrypted KV-cache mechanism, systems must repeatedly reprocess a growing prefix, yielding quadratic latency and memory growth.
(2) \textbf{Packing mismatch across phases:} the outer-based (token-batched) encodings used to maximize SIMD utilization in prefilling become sparse when only a single token is available in decoding, wasting slots and homomorphic work.
These challenges make naïve extensions of discriminative secure inference frameworks impractical for low-latency, long-context generation.

To overcome these challenges, we introduce \texttt{CryptoGen}, a scalable and low-latency system for secure neural network generation that intricately combines homomorphic encryption with traditional secret sharing. 
\texttt{CryptoGen} makes four key contributions:

\begin{itemize}
\item \textbf{Unified Secure Generation Framework:} 
\texttt{CryptoGen} establishes a unified yet modular cryptographic framework that explicitly separates the \emph{prefilling} and \emph{auto-regressive generation} phases while maintaining consistent cryptographic semantics across both. 
The framework decomposes Transformer inference into HE-based linear modules, MPC-based nonlinear primitives, and a lightweight orchestration layer that coordinates data exchange between them. 
This design provides a reusable, extensible library for secure generative inference and forms the foundation for cross-phase integration with encrypted Key–Value (KV) caches.

\item \textbf{Heterogeneous SIMD Encodings:} 
To accommodate the structural divergence between prefilling and decoding, \texttt{CryptoGen} adopts heterogeneous data encodings in encrypted computing. We use outer-based packing to maximize SIMD utilization across many tokens in prefilling, and switch to inner-based packing for single-token decoding to avoid sparse-slot waste. For linear layers, we leverage diagonal representations of weights to reduce rotation and permutation overhead. Together, these encodings bridge batched prefilling and token-by-token generation without forcing costly re-encoding of the entire context.

\item \textbf{Optimized Ciphertext–Ciphertext Computations:} 
We design new cipher–cipher multiplication algorithms tailored to encrypted attention in the autoregressive stage. 
Two complementary methods are proposed—an \emph{inner–inner} CT$\times$CT multiplication for generation-phase self-attention and an \emph{inner–outer} CT$\times$CT multiplication for cross-stage attention between generation and prefilling caches. 
Both algorithms incorporate logarithmic-depth rotations and element-duplication operations to realign ciphertext slots efficiently, reducing the original quadratic attention complexity to a linear form and achieving  significant speedup in encrypted attention and scalable performance for the long-sequence generation.

\item \textbf{Encrypted KV-Cache Management:} 
We introduce a fully encrypted KV-cache framework that supports noise-refresh and ciphertext-concatenation mechanisms for dynamic cache updates during generation. 
A collaborative HE–MPC refresh protocol securely restores ciphertext noise budgets, while a slot-aware concatenation scheme appends new encrypted keys and values into compact cache ciphertexts with minimal overhead. 
Together, these techniques maintain ciphertext freshness, memory efficiency, and continuity throughout long-sequence generation.

\end{itemize}

Together, these innovations bridge the gap between stateless encrypted inference and stateful autoregressive generation. We evaluate \texttt{CryptoGen} on generative Transformer models trained on WikiText-2, PTB, and LAMBADA, and show that for input lengths of 128--512 tokens, \texttt{CryptoGen} achieves $4.4\times$--$7.6\times$ lower per-token latency than prior secure inference systems designed for discriminative tasks, while maintaining near-linear latency and memory scaling as sequence length grows.

%% file: contents_usenix/2_background.tex
\subsection{Auto-regressive Transformer Generation}\label{subsec:generation}
In this paper, we focus on transformer-based generative models\cite{vaswani2017attention}.
Such generative tasks are typically divided into two stages: prefilling and auto-regressive generation.

\noindent\textbf{Prefilling.} 
At this stage, the model receives a complete prompt 
\(X=\{x_1,x_2,\ldots,x_m\}\), where each token embedding $x_i\in \mathbb{R}^{1\times d_1}$, $m$ denotes the length of the prompt and $d_1$ is the embedding dimension. The input prompt is processed through a stack of transformer layers, each containing a multi-head self-attention (MHSA) module followed by a feed-forward network (FFN)\cite{vaswani2017attention}.

For the $l$-th attention block with $h$ heads, the input $X\in \mathbb{R}^{m\times d_1}$ is linearly projected into query, key, and value representations for each head:
\[Q_h=XW_Q^{(h)}, K_h=XW_K^{(h)}, V_h=XW_V^{(h)},\]
where \(W_Q^{(h)},W_K^{(h)},W_V^{(h)}\in \mathbb{R}^{d_1\times d_2}\), and $d_2=d_1/H$ with $H$ being the number of attention heads.

Each head performs scaled dot-product attention:
\[
    O_h=\mathrm{Softmax}\left(\frac{Q_hK_h^{\top}}{\sqrt{d_2}}+M\right)V_h.
\]
where $M$ is a causal mask that enforces unidirectional dependencies in generative tasks\cite{vaswani2017attention}. The outputs from all heads are concatenated and linearly transformed:
\[O=\mathrm{Concat}(O_1,\ldots,O_H)W_{O},\] where $W_O\in \mathbb{R}^{d_1\times d_1}$ is the output projection matrix and $O\in \mathbb{R}^{m\times d_1}$ represents the attention output of the layer.

A position-wise feed-forward network is then applied to each token\cite{hendrycks2016gaussian}:\[
    FFN(x)=W_2GELU(W_1x+b_1)+b_2.
\]
Residual connections and layer normalization are applied after both the attention and feed-forward sublayers. 

Through these computations, the model encodes the entire prompt in parallel and produces key–value (KV) representations for each layer. These cached KV pairs serve as the contextual memory used during the subsequent auto-regressive decoding stage\cite{radford2019language}.

\noindent\textbf{Auto-regressive Generation.}
After the prefilling stage, the model enters the auto-regressive generation stage, in which new tokens are produced sequentially based on the cached context. In each decoding step $t>m$, the model takes as input the embedding of the previously generated token \(x_t\in \mathbb{R}^{1\times d_1}\) and projects it into \(Q^{(t)}_h,K^{(t)}_h,V^{(t)}_h\in \mathbb{R}^{1\times d_2}\) for each head .

Unlike the prefilling phase, the attention block computes attention only for the new query  $Q^{(t)}_h$ while reusing all previously cached keys and values: \[
O_h^{(t)} = \mathrm{Softmax}\left(\frac{Q_h^{(t)} [K_h^{(1)}, \ldots, K_h^{(t)}]^{\top}}{\sqrt{d_2}}\right) [V_h^{(1)}, \ldots, V_h^{(t)}].
\]
The results of all heads are concatenated, projected, and passed through the same feed-forward and normalization layers as in the prefilling stage to produce the hidden representation $h_t$ and the next-token prediction $P(x_{t+1}\mid x_{\le t})$.
% \lou{Prefilling}
% \lou{Auto-regressive Generation and it its connection with prefilling}

\subsection{KV Cache in Generation}\label{subsec:cache}
\begin{figure}[htbp]
    \centering
    \includegraphics[width=1\linewidth]{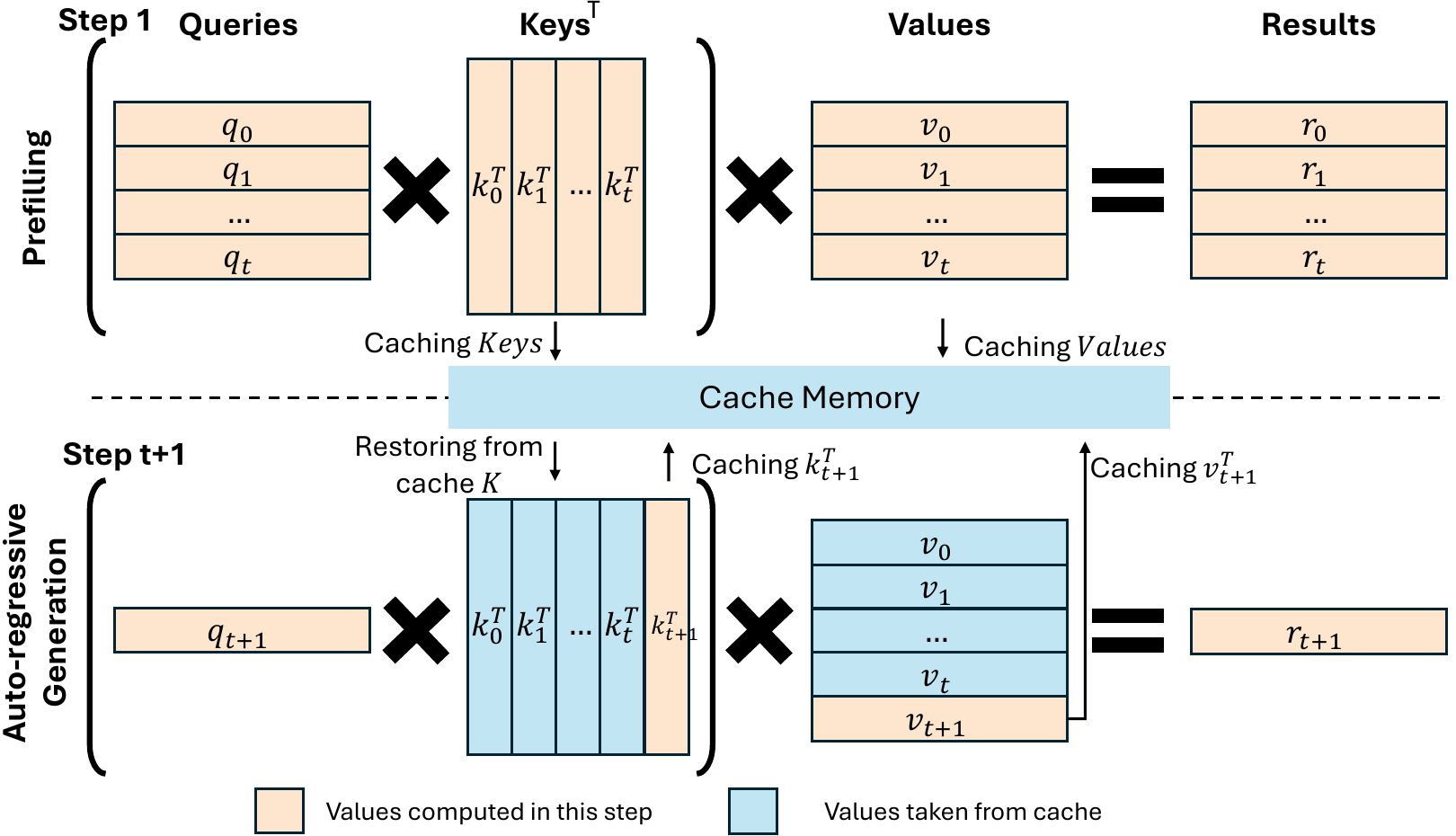}
    \captionsetup{skip=2pt}
    \caption{KV cache in auto-regressive Transformer generation.}
    \label{fig:kvcache}
\end{figure}
KV Cache is essential for efficient, scalable autoregressive generation.  This is because KV cache avoids redundant linear projections and repeated attention computation. Formally, by caching previously computed keys and values, the per-step attention cost can be reduced from quadratic $O(t^2)$ recomputation to a linear $O(t)$ incremental computation that only involves the new query. Figure~\ref{fig:kvcache} illustrates the standard key-value caching mechanism used in auto-regressive Transformer generation.
The output at step $t$ depends solely on the new query $q^{(t)}$ attending over all available cached keys and values, while previous queries $q^{(1)},\ldots,q^{(t-1)}$ do not participate in the current step. Therefore, the model can cache the previously computed keys and values from earlier tokens instead of recomputing them at every decoding step\cite{shoeybi2019megatron}. During each step, only the new projections $q^{(t)},k^{(t)},v^{(t)}$ are computed and the existing keys and values are reused from the cache.

% \lou{Definition, and its importance}

% \lou{How it works: cross prefilling and auto-regressive generation.}

\subsection{Cryptographic Primitives}
\textbf{Homomorphic Encryption.} Homomorphic Encryption (HE) is a powerful cryptographic technique that allows computations to be performed directly on encrypted data (i.e. ciphertexts) without first decrypting them\cite{gentry2009fully,gentry2013homomorphic}. The resulting ciphertext, when decrypted, matches the result of performing the same computations on the original plaintext data. HE schemes (particularly lattice-based FHE\cite{fan2012somewhat,brakerski2014leveled}) are well-suited for efficiently processing linear operations. This efficiency comes from two key properties. First, their underlying mathematical structure directly supports arithmetic operations such as additions and multiplications, which are the building blocks of matrix-vector products. Second, modern FHE schemes support "batching," allowing a single ciphertext to hold thousands of data values in SIMD (Single Instruction, Multiple Data) slots\cite{halevi2014algorithms}. This enables massive parallel computation, as a single homomorphic operation (e.g., a ciphertext-plaintext multiplication) can perform an entire layer's computation at once in a non-interactive manner on the server.

Recent work has advanced private inference and training over (fully) homomorphic encryption, including HE-friendly secure inference pipelines, private Transformer inference, and privacy-preserving NLP/vision models\cite{zhang2025cipherprune,zhang2024heprune,zheng2023primer,xue2023cryptotrain,lou2021hemet,feng2021cryptogru,lou2020autoprivacy}.
At the same time, systems and architecture efforts have improved the performance of FHE-style computation through better scheme support and hardware acceleration\cite{zhu2025dahe,kumar2025tfhe,deng2024trinity,zheng2024ofhe,zhang2023hebridge,zheng2023priml,zheng2022cryptolight,han2022coxhe,jiang2022matcha}.

Conversely, HE is fundamentally inefficient for non-linear functions (like GELU, ReLU, or Softmax)~\cite{lou2019she, lou2021safenet, yudha2024boostcom}. This is because HE operations are inherently polynomial. Non-linear functions must be \textit{approximated} by low-degree polynomials (e.g., Taylor series). This approach has two major drawbacks: (1) it introduces precision errors that can harm model accuracy, and (2) evaluating these polynomials requires multiple, slow ciphertext-ciphertext multiplications. Each such multiplication significantly increases the computational "noise" in the ciphertext, which quickly grows to unmanageable levels, requiring an extremely costly "bootstrapping" procedure (a noise reset) that makes deep computations impractical.

\textbf{Multi-Party Computation.}
Multi-Party Computation (MPC) enables a set of mutually distrusting parties (e.g., a client with a private input and a server with a private model) to jointly compute a common function without any party revealing their private inputs\cite{yao1982protocols}. MPC protocols typically rely on secret-sharing mechanisms. MPC is efficient for non-linear functions. Unlike HE, MPC can execute operations like comparisons (the basis of ReLU) or secure exponentiation (for Softmax) \textit{exactly} and relatively cheaply using specialized sub-protocols (like garbled circuits or secure comparisons) that require only a few rounds of low-bandwidth communication. This avoids the approximation errors and extreme "bootstrapping" costs associated with HE\cite{li2022mpcformer}.

However, MPC is not suitable for large-scale linear operations found in Transformers. Performing a matrix-vector multiplication in MPC would require the client and server to exchange a large volume of intermediate shares over the network for every single layer. This high communication overhead and sensitivity to network latency make it significantly slower for linear algebra than HE, which, as mentioned, performs these operations non-interactively on the server.

\textbf{Hybrid HE and MPC Frameworks.}
Hybrid frameworks are widely used in secure inference to combine the respective strengths of HE and MPC\cite{juvekar2018gazelle,pang2024bolt,hao2022iron,lu2023bumblebee}. In a typical hybrid design, the server executes compute-intensive and SIMD-friendly linear operations in the HE domain. When a non-linear function is encountered, the HE ciphertext is converted into MPC secret shares via a secure conversion protocol. The client and server then jointly execute an MPC protocol to evaluate the function. Finally, the resulting shares are converted back into an HE ciphertext for subsequent linear computations.

Therefore, CryptoGen also adopted this hybrid framework.

\subsection{Data Encodings in Encrypted Computing}

% \begin{figure}[!t]
\begin{figure}[htbp]
    \centering
    \includegraphics[width=1\linewidth]{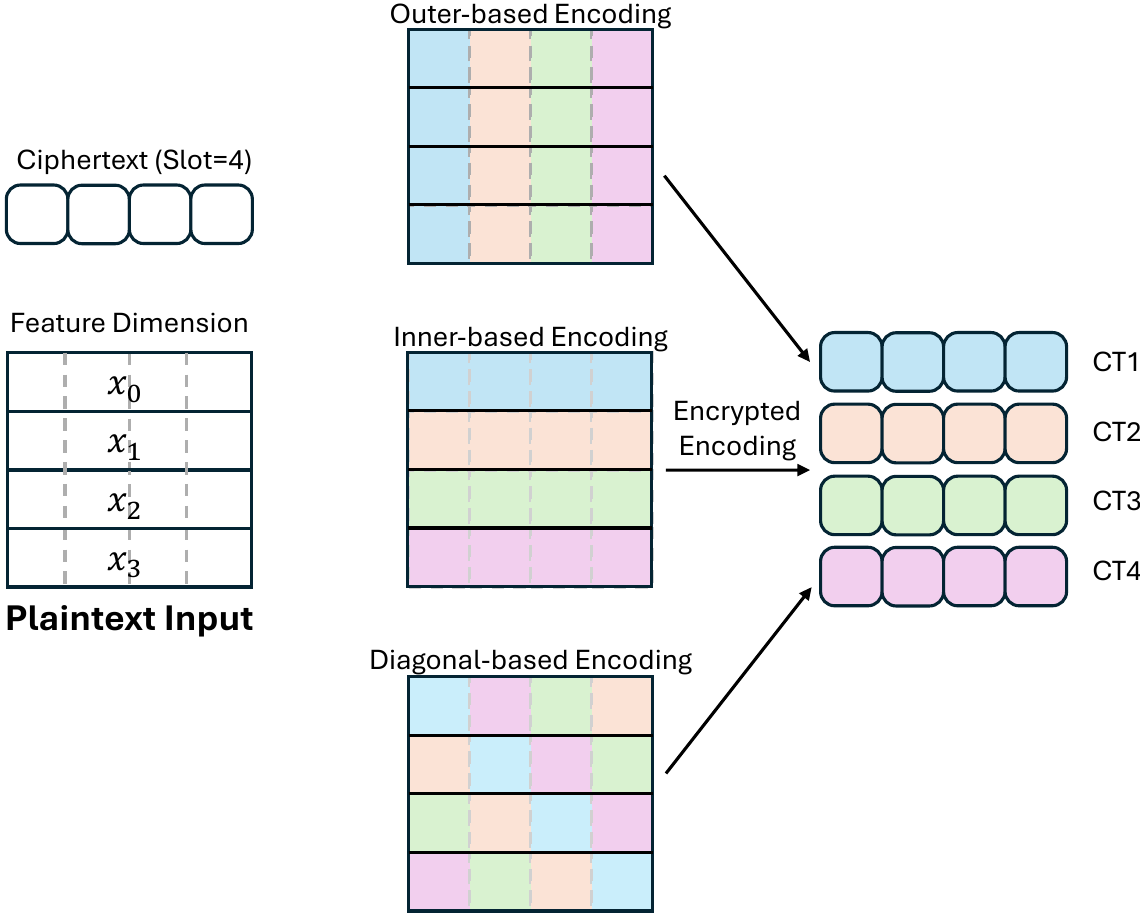}
    \captionsetup{skip=2pt}
    \caption{Three ciphertext packing strategies in encrypted computing: (a) outer-based, (b) inner-based, and (c) diagonal encoding. Different packings expose different forms of SIMD parallelism and lead to different trade-offs between the number of ciphertexts, cross-ciphertext reductions, and rotation overhead.}
    \label{fig:packing}
\end{figure}

\label{subsec:encoding}

In HE, a ciphertext can pack a vector of plaintext values into SIMD slots, enabling one homomorphic operation to be applied to many elements in parallel\cite{halevi2014algorithms}. When implementing Transformer inference over encrypted data, the choice of packing (i.e., how we map a matrix or vector into ciphertexts) directly determines the amount of available parallelism and the cost of HE primitives such as ciphertext rotations and slot-wise aggregations.

We summarize three commonly used encodings for encrypted linear algebra, as illustrated in Figure~\ref{fig:packing}.

\noindent\textbf{Outer-based  encoding.}
Given a matrix $A\in\mathbb{R}^{m\times n}$, outer-based encoding packs each \emph{column} $A_{:,j}$ into one ciphertext. We call it \emph{outer-based} because it naturally matches an outer-product style matrix multiplication: when computing $y=Ax$, each ciphertext-column can be scaled by the corresponding scalar (or packed) entry $x_j$ and then accumulated across columns. This encoding exposes high parallelism across columns, but it typically requires aggregating (adding) many ciphertexts to form the final result.

\noindent\textbf{Inner-based encoding.}
Inner-based encoding packs each \emph{row} $A_{i,:}$ into one ciphertext. This layout aligns with inner-product computation during matrix multiplication: the $i$-th output element is the inner product $\langle A_{i,:},x\rangle$, which can be realized by an element-wise multiplication between the packed row and a packed input vector, followed by a slot-summation (usually implemented via a sequence of rotations and additions). Compared to outer-based encoding, inner-based encoding reduces cross-ciphertext accumulation but often increases within-ciphertext rotation overhead for the reduction.

\noindent\textbf{Diagonal encoding.}
Diagonal encoding packs a \emph{row-first diagonal} of $A$ into one ciphertext. Concretely, the $k$-th diagonal ciphertext contains elements $A_{i,(i+k)\bmod n}$ for $i\in\{1,\ldots,m\}$ (with appropriate padding when dimensions do not match). This encoding is widely used to implement $y=Ax$ with a small number of ciphertexts: multiplying each diagonal ciphertext with a (rotated) version of the packed vector $x$ and then summing over diagonals yields $y$. Diagonal encoding often reduces the number of ciphertexts but relies heavily on rotations to align slots.

%% file: contents_usenix/threat_model.tex
% We consider a standard two-party client--server setting under the semi-honest (honest-but-curious) adversary model. 
% The server follows the prescribed protocol but may attempt to infer sensitive information from intermediate states.
% The client holds the decryption key and does not collude with the server.
% Our goal is to protect the confidentiality of user inputs and generated tokens, including the KV cache.
% We do not consider malicious adversaries, side-channel attacks, or integrity violations in this work.

We consider a client-server setting for outsourced inference under the semi-honest (honest-but-curious) model, a fundamental and widely used assumption across secure computation and privacy-preserving ML\cite{juvekar2018gazelle,pang2024bolt,moon2024thor}: the server follows the protocol but may try to infer information from what it observes, while the client holds the secret key material and does not collude with the server. Our goal is confidentiality of the client prompt and generated outputs; in particular, the KV cache is maintained and reused only in encrypted form so that it does not reveal token content. We do not address malicious deviations, side-channel leakage, or integrity/verifiability in this work, as these threats are largely orthogonal to our focus and can be incorporated via established complementary techniques\cite{pang2024bolt,moon2024thor}.

%% file: contents_usenix/3_motivation.tex
\subsection{Prior Work: Secure Inference for Discriminative Transformers}
\label{subsec:motivation_discriminative}

Most prior systems for privacy-preserving Transformer inference target \emph{discriminative} (encoder-style) workloads, where the entire input sequence is available before inference and the computation can be executed in a single batched pass. Table~1 summarizes representative systems, including Gazelle\cite{juvekar2018gazelle}, IRON\cite{hao2022iron}, BOLT\cite{pang2024bolt}, and THOR\cite{moon2024thor}. Although some of these works evaluate Transformer \emph{generative} models, their protocols still focus on discriminative settings (e.g., classification) or on producing a \emph{single} next-token output---which is closer to the prefilling stage than to token-by-token decoding.

Among these systems, BOLT\cite{pang2024bolt} is the state-of-the-art baseline in our comparison. Its performance largely comes from a carefully chosen \emph{outer--diagonal} encoding strategy for linear layers: it packs the input activation matrix $X$ using an outer-based (column-wise) encoding, and represents the weight matrix $W$ in a diagonal form (i.e., row-first diagonals). This enables fast ciphertext--ciphertext matrix multiplication with fewer rotations and fewer homomorphic operations than inner-product based formulations, while avoiding slot waste when the token dimension is large (as in prefilling). In short, BOLT achieves high throughput by maximizing SIMD utilization over many tokens and minimizing rotation-heavy reductions.

\subsection{Why Prior Works Do Not Extend to Auto-regressive Generation}
\label{subsec:motivation_bolt_limit}
BOLT's outer--diagonal design is tailored to prefilling-style computation, where the token dimension is large and can fully occupy ciphertext slots. In auto-regressive decoding, however, the model consumes \emph{one} new token at a time and must attend to an ever-growing history stored in the KV cache. This breaks the core assumption behind outer-based packing: that SIMD slots can be filled with many independent tokens.
\begin{figure}[htbp]
    \centering
    \includegraphics[width=1\linewidth]{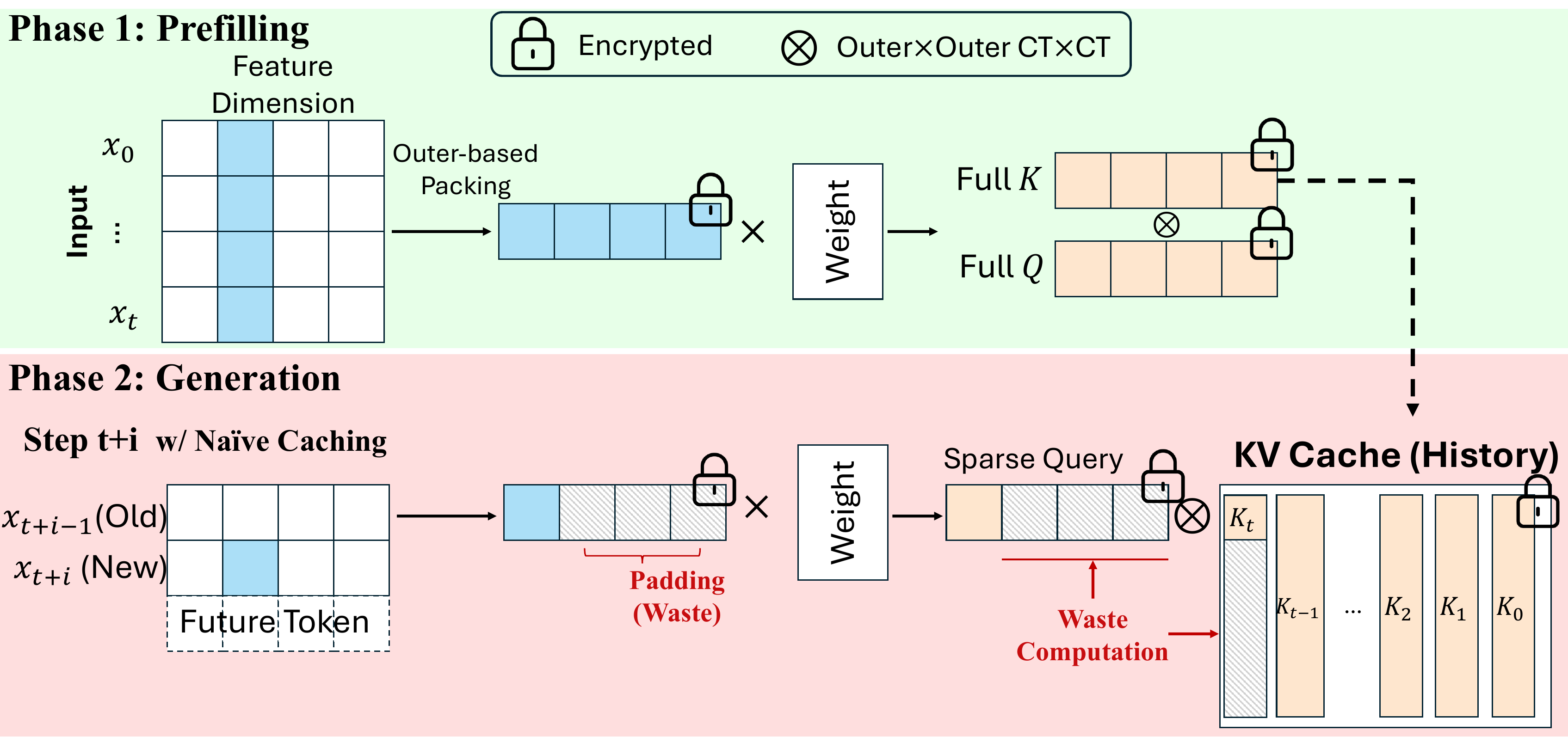}
    \captionsetup{skip=2pt}
    \caption{Why outer-based encodings (including BOLT's outer--diagonal design) do not fit token-by-token decoding with an encrypted KV cache: after prefilling, each step introduces only one new token, leading to sparse ciphertext utilization or redundant recomputation.}
    \vspace{-6pt}
    \label{fig:motivation}
\end{figure}
Figure~\ref{fig:motivation} illustrates this mismatch. During prefilling (Phase~1), outer-based packing is ideal because the batch of tokens densely fills SIMD slots. After switching to decoding, the new token at step $t+i$ occupies only a tiny fraction of slots, leaving most slots as padding (Phase~2). With an encrypted KV cache, the system either (i) performs attention and linear layers on sparse ciphertexts (wasting work on padding), or (ii) recomputes past tokens to artificially refill slots (wasting work on redundant computation).

Crucially, if the KV cache cannot be reused efficiently, the system must repeatedly re-encode and re-process a growing prefix, leading to latency and memory costs that grow \emph{quadratically} with sequence length rather than linearly.

\subsection{Motivation: Secure Generation with Encrypted KV Cache Reuse}
\label{subsec:motivation_goal}
\begin{figure}[htbp]
    \centering
    \includegraphics[width=1\linewidth]{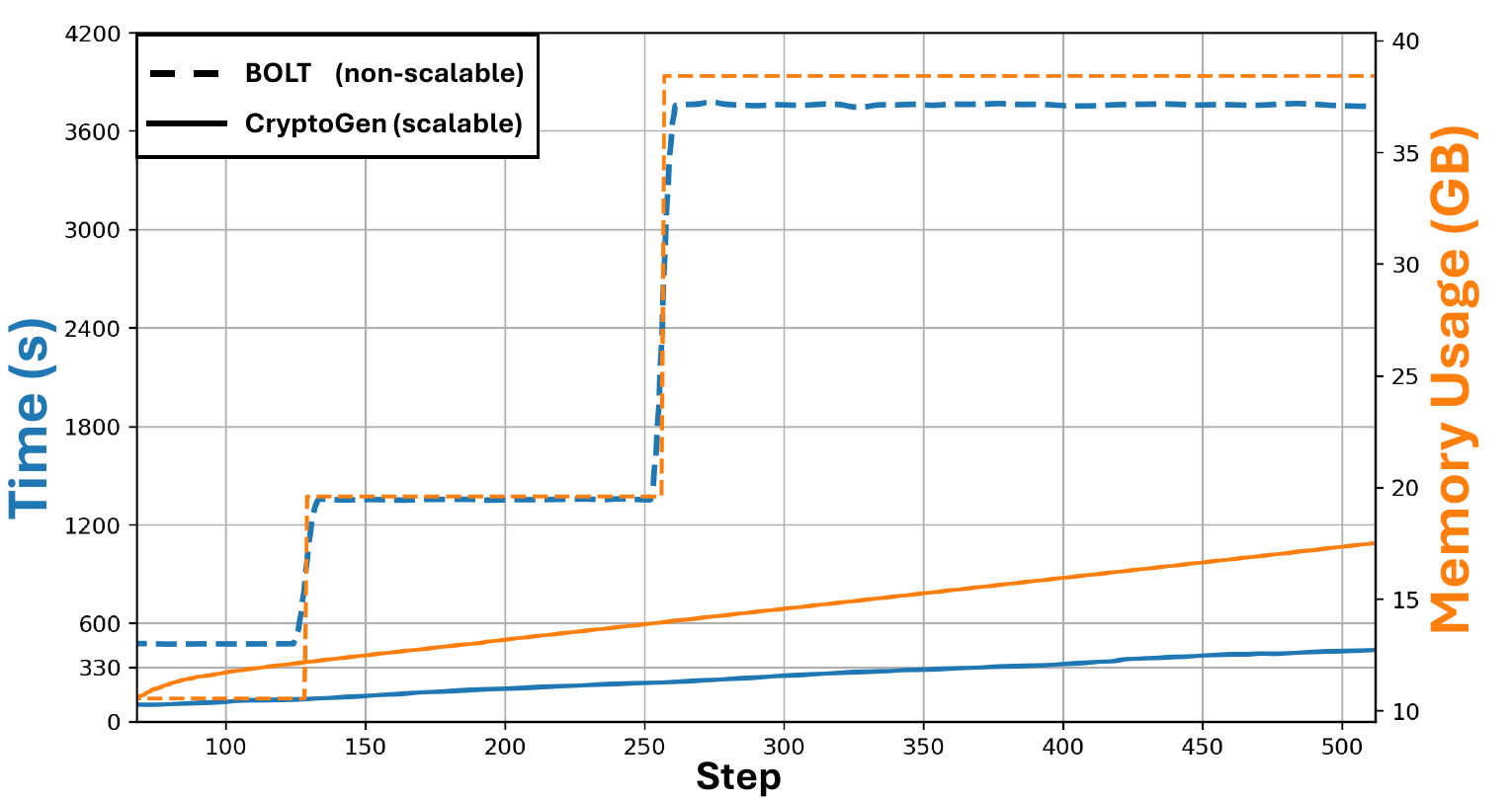}
    \captionsetup{skip=2pt}
    \caption{Scalability comparison of secure auto-regressive generation. The latency of discriminative inference framework(BOLT) is unscalable due to the lack of KV cache reuse. CryptoGen maintains near-linear scaling with sequence lenght by using KV cache. Refer to Section \ref{subsec:end2end} for detailed end-to-end performance analysis.
    % Scalability of secure generation with and without effective KV-cache reuse. Plaintext decoding reuses the KV cache and scales linearly with sequence length. Outer-based approaches that cannot efficiently reuse an encrypted KV cache incur quadratic-like latency growth (due to sparse padding) or increasing per-step cost (due to recomputation).
    }
    \label{fig:scalability}
    \vspace{-0.2in}
\end{figure}
Figure~\ref{fig:scalability} highlights the core gap in prior work: existing HE-based systems are optimized for batched, discriminative inference, but do not provide an efficient mechanism to \emph{reuse} and \emph{update} the KV cache throughout token-by-token decoding. Without such support, secure generation becomes fundamentally unscalable in both latency and memory.

This motivates CryptoGen: we aim to build a secure inference framework specialized for auto-regressive generation that (i) maintains an encrypted KV cache across steps, (ii) supports efficient per-token incremental computation, and (iii) achieves near-linear scaling with sequence length, analogous to plaintext decoding. In the remainder of the paper, we present the design of CryptoGen and show how its encoding choices and linear-algebra kernels are tailored to stateful decoding rather than one-shot prefilling.

%% file: contents_usenix/5_method.tex
\subsection{CryptoGen Framework}\label{sec:system_overview} %\ql{put it into the end of the section and introduce the proposed encodings first, which would highlight the novelty;}
% \begin{figure}[!t]
\begin{figure}[htbp]
    \centering
    \includegraphics[width=1\linewidth]{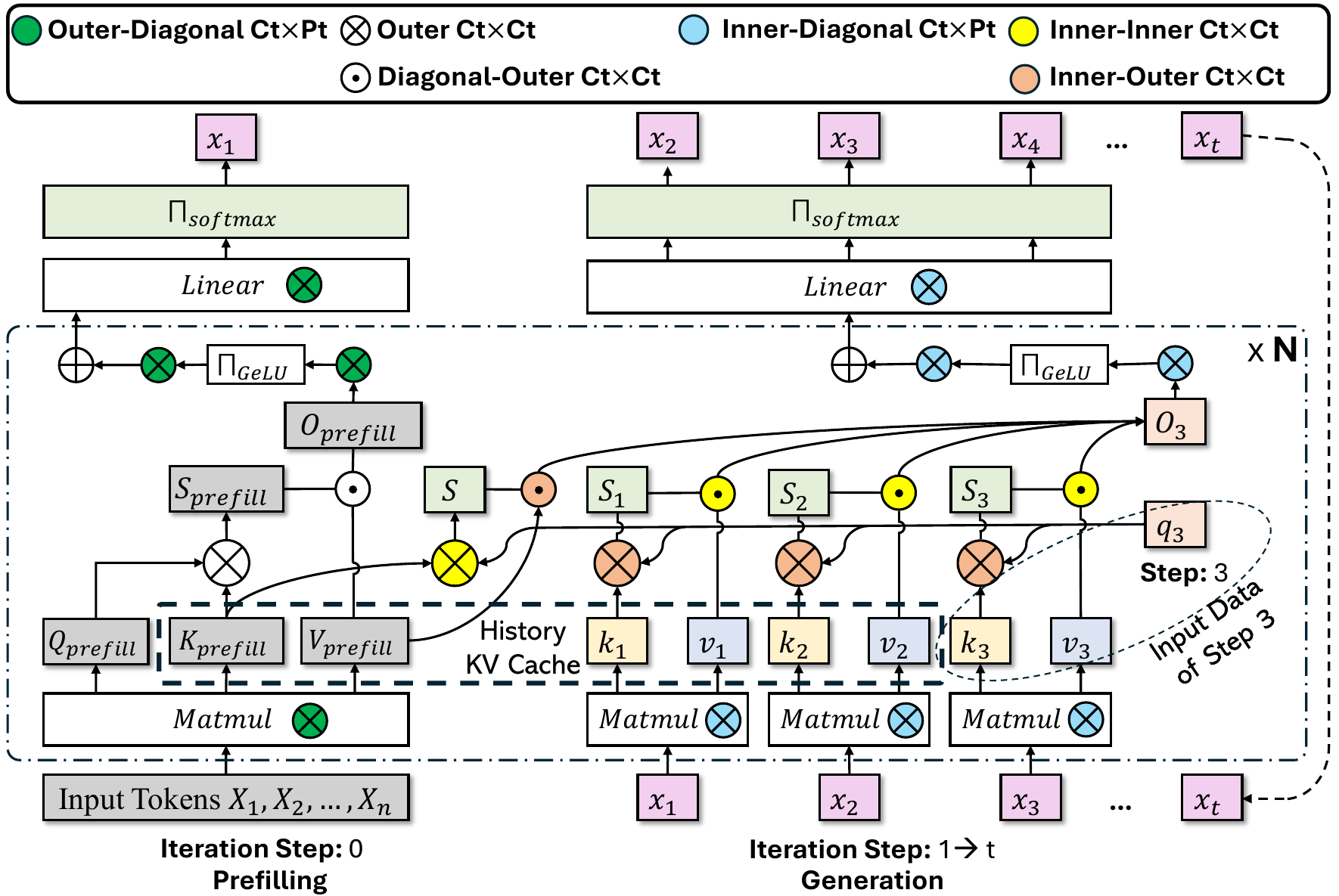}
    \captionsetup{skip=2pt}
    \caption{
CryptoGen inference workflow. It combines (i) prefilling with outer-diagonal CT$\times$PT matrix multiplication (CPMM) and (ii) autoregressive decoding with inner-diagonal CT$\times$PT vector-matrix multiplication (CPVM), bridged by a heterogeneous encrypted KV cache and MPC-based nonlinear primitives.}
    \label{fig:workflow}
\end{figure}
Current HE-based Transformer inference frameworks are predominantly designed for discriminative tasks, operating under a stateless execution model. They fail to accommodate autoregressive generation, where computation shifts from batch-parallel prefilling to token-sequential decoding and must reuse a growing Key-Value (KV) cache.

\textit{CryptoGen} addresses this gap with a \emph{heterogeneous-encoding} framework that aligns ciphertext layout with the available parallelism in each phase, enabling encrypted KV-cache reuse and fast end-to-end decoding. Figure~\ref{fig:workflow} summarizes the process and the corresponding primitive types.

\noindent\textbf{Process Overview (Figure~\ref{fig:workflow}).}
CryptoGen executes Transformer inference as a pipeline that alternates between CT$\times$PT linear layers (ciphertext activations, plaintext weights) and CT$\times$CT attention operations (both operands encrypted), while maintaining a persistent, encrypted KV cache:

\textbf{Prefilling phase (batch-parallel).}
Given a prompt $\{X_1,\ldots,X_m\}$, CryptoGen packs tokens with \emph{outer-diagonal encoding} and evaluates CT$\times$PT matrix multiplication (outer-diagonal CPMM) for the linear projections (e.g., $Q_{prefill},K_{prefill},V_{prefill}$). Attention during prefilling is evaluated using \emph{outer-outer} CT$\times$CT operations over the outer-packed activations.
The resulting $K_{prefill}$ and $V_{prefill}$ are stored as the \emph{outer-packed} segment of the encrypted KV cache.

\textbf{Decoding phase (token-sequential).}
At each decoding step $t$, the input collapses to a single token $X_t$. CryptoGen switches to \emph{inner-diagonal encoding} and evaluates CT$\times$PT vector-matrix multiplication (inner-diagonal CPVM) to compute $(q_t,k_t,v_t)$ without padded slots.
To consume the heterogeneous KV cache, attention is computed via encoding-aware CT$\times$CT kernels:
(i) \emph{inner-inner} CT$\times$CT between inner-packed operands, and
(ii) \emph{inner-outer} CT$\times$CT between an inner-packed operand and an outer-packed operand.
In addition, CryptoGen uses a \emph{diagonal-outer} CT$\times$CT multiplication when a diagonal-aligned operand must interact with an outer-packed cache block.
The newly generated $(k_t,v_t)$ are appended as the \emph{inner-packed} segment of the KV cache.

\textbf{KV-cache management (stateful, long-context).}
Heterogeneous encoding makes KV reuse possible, but it also requires careful ciphertext management: CryptoGen stores outer-packed and inner-packed cache segments compactly and reduces noise overhead via refresh/compaction mechanisms (Section~4.4).

\noindent\textbf{Module Decomposition.}
Accordingly, CryptoGen consists of: (1) \textbf{heterogeneous encoding and CT$\times$PT linear modules} (Section~\ref{ss:CT-PT}), which explain why our outer-/inner-diagonal encodings improve CT$\times$PT efficiency over prior encodings; (2) \textbf{heterogeneous-encoding CT$\times$CT kernels} for attention (Section~\ref{sec:ctct}), including outer-outer, inner-inner, inner-outer, and diagonal-outer interactions; and (3) \textbf{encrypted KV-cache management} (Section~\ref{ss:KVM}) to store heterogeneous cache states compactly and reduce noise/communication overhead.

\subsection{Heterogeneous Encoding based CT-PT Multiplication}\label{ss:CT-PT}
\noindent\textbf{Design principles.}
CryptoGen’s CT$\times$PT linear layers are the dominant building blocks for the $Q/K/V$ projections and feed-forward networks, but their efficiency is highly sensitive to ciphertext packing. Our design follows three principles.
(1) \emph{Match packing to input cardinality:} use batch-parallel packing during prefilling and single-token packing during decoding to avoid padded slots.
(2) \emph{Minimize per-step work during decoding:} the CT$\times$PT cost for generating the next token should be independent of the prompt length $m$.
(3) \emph{Preserve KV-cache reuse:} the packing choices must be compatible with later CT$\times$CT attention over a persistent (and heterogeneous) encrypted KV cache.

\begin{table*}[htbp]
\centering
\footnotesize
\resizebox{\textwidth}{!}{
\begin{tabular}{|c|
c|c|c|
c|c|c|
c|c|c|}
\hline
\textbf{Methods} &
\multicolumn{3}{c|}{\textbf{Mult}} &
\multicolumn{3}{c|}{\textbf{Rot}} &
\multicolumn{3}{c|}{\textbf{Ct}} \\ \hline
\textbf{Stage} &
\textbf{Prefill}& \textbf{Gen}& \textbf{Total} &
\textbf{Prefill}& \textbf{Gen}& \textbf{Total} &
\textbf{Prefill}& \textbf{Gen}& \textbf{Total} \\ \hline

\textbf{Gazelle}*&
$O(md_1)$ & $O(md_1\cdot k)$& &
$O(md_1)$ & $O(md_1\cdot k)$& &
$O(\tfrac{md_1}{d_2})$ & $O(\tfrac{md_1}{d_2}\cdot k)$& \\ 
& 98304 & 491520& 589824&
96768 & 483840& 580608&
1664 & 8320& 9984\\ \hline

\textbf{IRON}*&
$O(\tfrac{md_1d_2}{n})$ & $O(\tfrac{md_1d_2}{n}\cdot k)$& &
0 & 0 & 0 &
$O(\sqrt{\tfrac{md_1d_2}{n}})$ & $O(\sqrt{\tfrac{md_1d_2}{n}}\cdot k)$& \\ 
& 768 & 3840& 4608&
 0&  0&  0&
56 & 280& 336\\ \hline

\textbf{BOLT}*&
$O(\tfrac{md_1d_2}{n})$ & $O(\tfrac{md_1d_2}{n}\cdot k)$& &
$O(\sqrt{\tfrac{m^2d_1^2d_2}{n^2}})$& $O( \sqrt{\tfrac{m^2d_1^2d_2}{n^2}}\cdot k)$& &
$O(\tfrac{m(d_1+d_2)}{n})$ & $O(\tfrac{m(d_1+d_2)}{n}\cdot k)$& \\
& 768 & 3840& 4608&
43& 215& 258&
12& 60& 72\\ \hline

\textbf{ThOR}*&
$O(\tfrac{md_1d_2}{n})$ & $O(\tfrac{md_1d_2}{n}\cdot k)$& &
$O(d_2+\tfrac{md_1}{n})$ & $O((d_2+\tfrac{md_1}{n})\cdot k)$& &
$O(\tfrac{md_1}{n})$ & $O(\tfrac{md_1}{n}\cdot k)$& \\
& 9908 & 49540& 59448&
282 & 1410& 1692&
13 & 65& 78\\ \hline

\textbf{CryptoGen} &
$O(\tfrac{md_1d_2}{n})$ & $O(\tfrac{d_1d_2}{n}\cdot k)$& &
$O(\sqrt{\tfrac{m^2d_1^2d_2}{n^2}})$& $O(\log d_1\cdot k)$& &
$O(\tfrac{m(d_1+d_2)}{n})$ & $O(\lceil\frac{d_1}{n}\rceil\cdot k)$& \\
& 768 & 64$\times$5& 1088&
43& 25& 68&
12& 1$\times$5& 17\\ \hline
\end{tabular}
}
\captionsetup{skip=2pt}
\caption{CT$\times$PT complexity comparison under different packing strategies. We report the number of ciphertext-plaintext multiplications (Mult), homomorphic rotations (Rot), and ciphertexts (Ct) in prefilling and autoregressive decoding (Gen, $k$ generated tokens). Parameters: $m=128, d_1=768, d_2=64, n=8192$.\\
*Gazelle, IRON, BOLT, and ThOR are not designed for autoregressive decoding with KV cache reuse; they pre-allocate empty slots for future tokens and append generated tokens to the padded prefill sequence to emulate autoregression.}
\label{tab:complexity}
\end{table*}
\noindent\textbf{Why heterogeneous encoding for CT$\times$PT?}
Table~\ref{tab:complexity} previews the key consequence of our design: by switching packing layouts across phases, CryptoGen reduces the per-step CT$\times$PT cost during decoding from being proportional to the prompt length $m$ to being independent of $m$.
As highlighted in Figure~\ref{fig:workflow}, Transformer inference alternates between CT$\times$PT projections (ciphertext activations, plaintext weights) and CT$\times$CT attention over a persistent KV cache. A single packing scheme cannot simultaneously (i) exploit token-level SIMD parallelism in prefilling and (ii) avoid padded slots when decoding one token at a time.

\begin{figure}[htbp]
    \centering
    \includegraphics[width=1\linewidth]{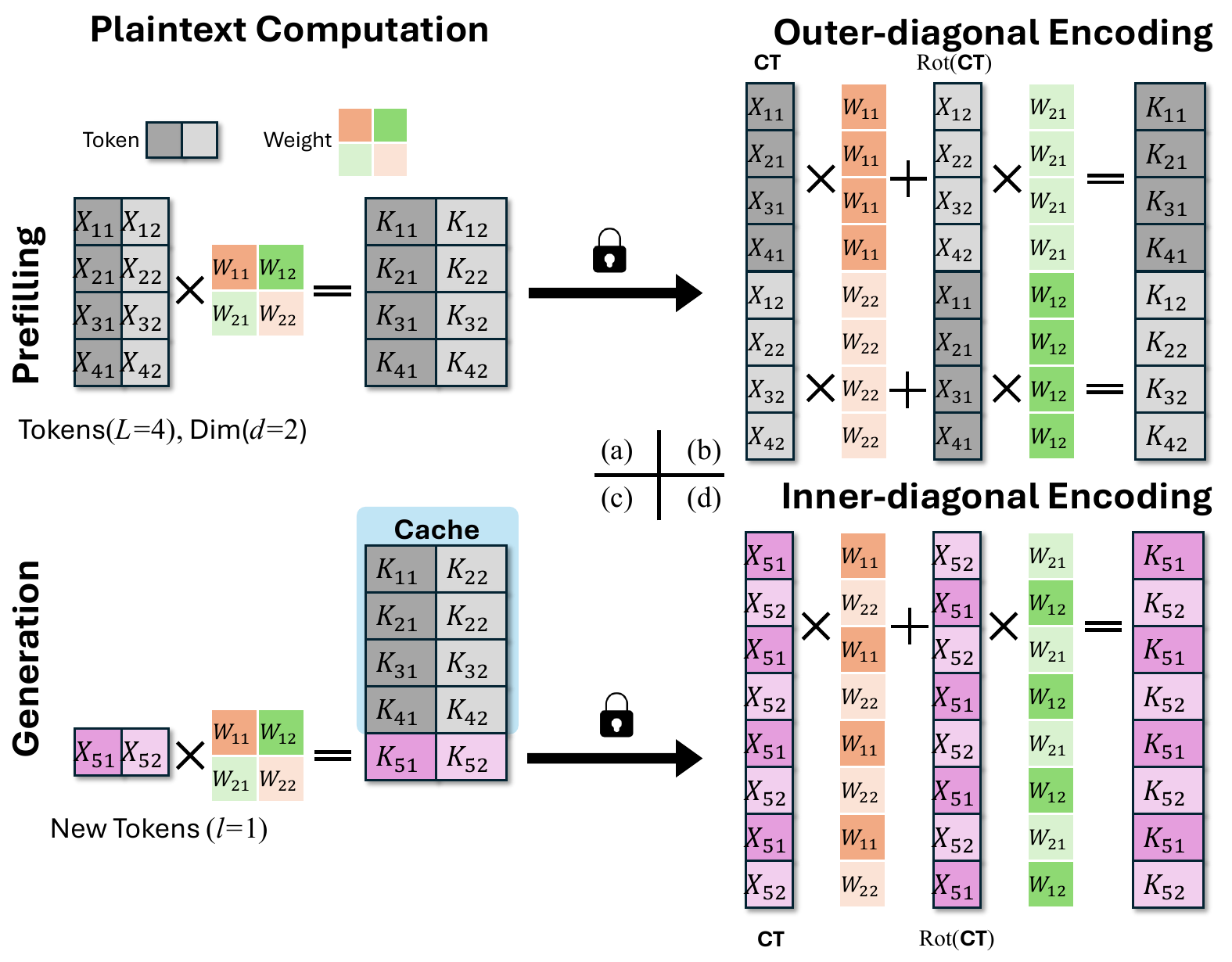}
    \captionsetup{skip=2pt}
    \caption{
Heterogeneous encoding layouts for CT$\times$PT linear layers.
(a, b) outer-diagonal encoding packs a batch of tokens ($X_1 \dots X_4$) across slots for batch-parallel prefilling (outer-diagonal CPMM).
(c, d) inner-diagonal encoding packs one token ($X_5$) across slots for token-sequential decoding (inner-diagonal CPVM).
}
    \label{fig:encode}
\end{figure}

\noindent\textbf{Prefilling implementation (outer-diagonal CPMM).}
During prefilling, the input is a multi-token prompt $X \in \mathbb{Z}_p^{m\times d_1}$. CryptoGen uses \emph{outer-diagonal encoding} to pack a batch of tokens across ciphertext slots (Figure~\ref{fig:encode}(a,b)) and evaluates CT$\times$PT \emph{ciphertext-plaintext matrix multiplication} (CPMM). This realizes batch-parallel $Q/K/V$ projections with high SIMD utilization and produces outer-packed $K_{prefill}$ and $V_{prefill}$ that can be reused directly by attention.

\noindent\textbf{Decoding implementation (inner-diagonal CPVM).}
During autoregressive decoding, the input collapses to a single token $x_t \in \mathbb{Z}_p^{1\times d_1}$. CryptoGen switches to \emph{inner-diagonal encoding} to pack one token across slots (Figure~\ref{fig:encode}(c,d)) and evaluates CT$\times$PT \emph{ciphertext-plaintext vector-matrix multiplication} (CPVM), mapping feature dimensions across slots to avoid padding. This makes the per-step CT$\times$PT cost independent of $m$ while producing inner-packed $(q_t,k_t,v_t)$ that are compatible with heterogeneous CT$\times$CT attention kernels in Section~4.3.

% \begin{figure}[!h]

\noindent\paragraph{Complexity Comparisons of Heterogeneous Encoding-based CPMM and priors.} In
Table~\ref{tab:complexity}, we analyze the computational complexity of ciphertext-plaintext linear layers under CryptoGen’s heterogeneous encoding strategy, with a particular focus on the per-step cost during auto-regressive generation. In this regime, scalability critically requires that computation be independent of the prefilling batch size $m$.

Existing HE-based Transformer inference systems—including Gazelle, IRON, ThOR, and BOLT—adopt batch-oriented encoding layouts originally designed for stateless, discriminative workloads. While their packing strategies differ (inner-based encoding, block-based encoding, diagonal-based encoding, and outer-based encoding, respectively), all of these methods preserve a batch-parallel ciphertext layout during generation. As a result, the single newly generated token must be padded into a batch of size $m$, causing the per-step generation cost to remain proportional to the prompt length.

In contrast, CryptoGen decouples generation complexity from the prefilling batch size by enforcing heterogeneous encoding across inference stages. Outer-diagonal CPMM is retained during prefilling to exploit token-level parallelism, while inner-diagonal CPVM is activated during decoding to match the single-token input cardinality. Consequently, the per-step ciphertext-plaintext cost during generation is reduced from $O\!\left(\tfrac{m d_1 d_2}{n}\right)$ to $O\!\left(\tfrac{d_1 d_2}{n}\right)$, eliminating the dominant source of inefficiency present in batch-oriented encodings.

Furthermore, heterogeneous encoding enables more efficient accumulation within each generation step. Whereas prior approaches require $O(d_1)$ or $O(m)$ homomorphic rotations due to linear-depth slot aggregation or padded batch processing, CryptoGen performs local aggregation over compact inner-based encoding ciphertexts, reducing the rotation complexity to $O(\log d_1)$. The ciphertext footprint per decoding step is likewise bounded by $O(\lceil d_1/n \rceil)$, independent of the prompt length.

These complexity reductions are summarized quantitatively in Table~\ref{tab:complexity}. By aligning ciphertext layout with the cardinality of the input at each inference stage, heterogeneous encoding allows CryptoGen to preserve high throughput during prefilling while achieving scalable, low-latency decoding during auto-regressive generation.

While heterogeneous encoding reshapes the complexity of linear projections, attention computation further requires resolving interactions over a heterogeneous encrypted KV cache, which we address in the next section.

\subsection{Auto-Regressive Ciphertext-Ciphertext Multiplication (ARCC)}\label{sec:ctct}
In autoregressive decoding, the attention mechanism computes the score vector $s = qK^T$ and the context vector $o = aV$, where $a = \mathrm{Softmax}(s)$. Unlike CT$\times$PT linear layers, these operations are inherently stateful because they must interact with the encrypted KV cache accumulated from the prompt and previously generated tokens.

Due to CryptoGen’s heterogeneous encoding strategy, the KV cache is \emph{heterogeneous}: the prefilling history is stored in an outer-packed layout, while the generated history is stored in an inner-packed layout. Consequently, standard homogeneous CT$\times$CT matrix multiplication protocols are no longer directly applicable. Forcing a single unified packing would either destroy decoding efficiency (by reintroducing padded slots) or eliminate KV reuse (by requiring re-encoding at every step).

At each step, the inner-packed query must therefore interact efficiently with both cache segments. We address this with \emph{Autoregressive Ciphertext-Ciphertext Multiplication (ARCC)}, a family of encoding-aware CT$\times$CT kernels that implement incremental attention over heterogeneous KV caches.
\begin{figure}[htbp]
    \centering
    \includegraphics[width=0.5\textwidth]{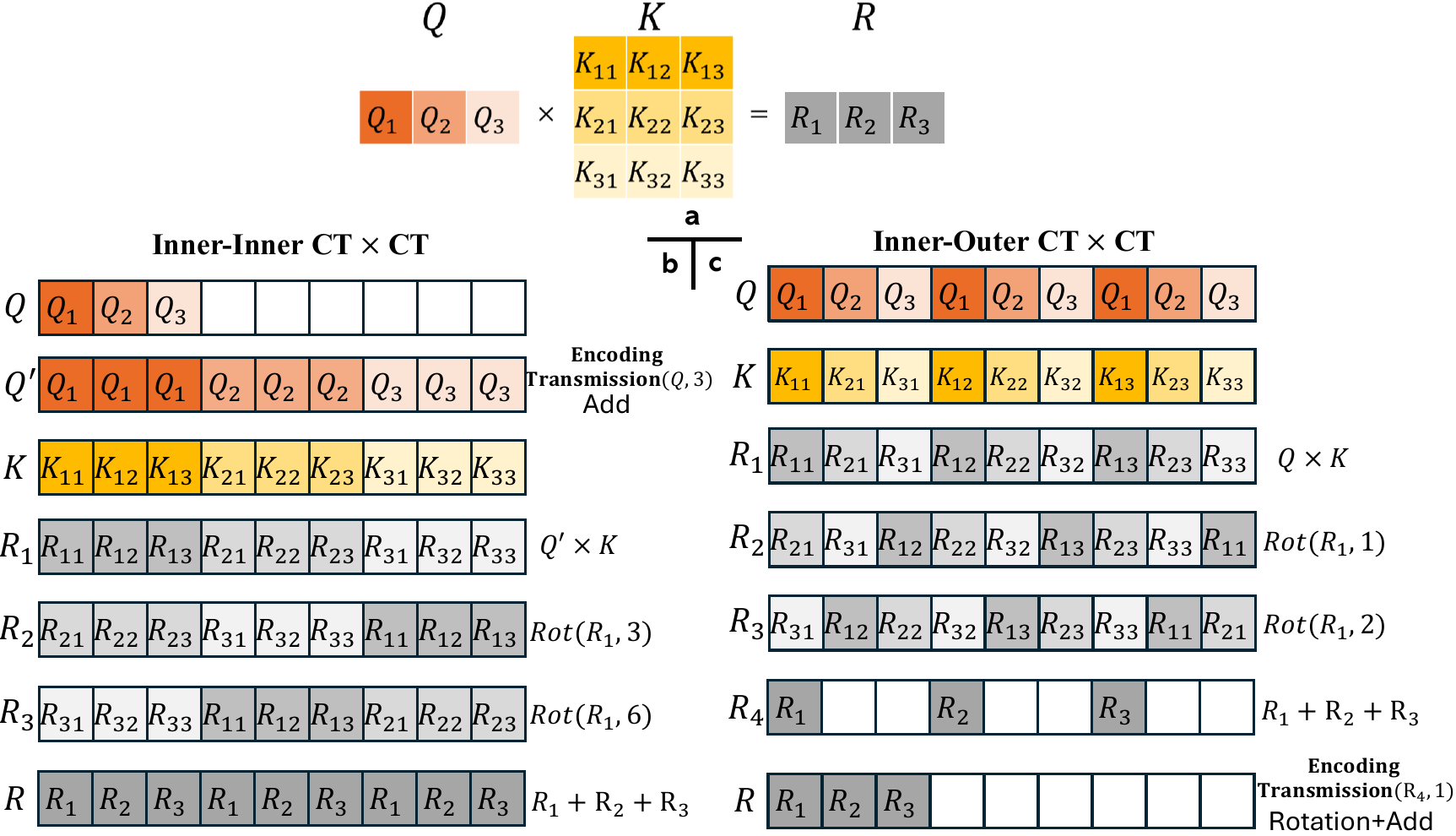} 
    \captionsetup{skip=2pt}
    \caption{Autoregressive Ciphertext-Ciphertext (ARCC) kernels for attention over a heterogeneous KV cache: (a) plaintext $qK^T$ for reference; (b) ARCC inner-inner mode (query with the transposed outer-packed prefilling cache); (c) ARCC inner-outer mode (query with the transposed inner-packed generated cache).}
    \vspace{-7pt}
    \label{fig:ctct}
\end{figure}

\noindent\textbf{ARCC Inner–Inner Mode.}
From an algebraic perspective, this kernel handles the multiplication of two operands that both possess an inner-based packing format. In the autoregressive phase, this logic is applied to compute the attention scores against the prefilling history $s_{prefill} = q K_{prefill}^T$ and to aggregate the values for the generated history $o_{auto} = a V_{auto}$. For $s_{prefill}$, although the prefilling cache $K_{prefill}$ is physically stored in the outer-based packing layout, the attention operation targets its transpose $K_{prefill}^T$. Algebraically, the transpose of an outer-based packing matrix behaves as an inner-based packing format, matching the inner-packed query $q$. Thus, we treat this as an inner-inner interaction.
\begin{figure}[htbp]
    \centering
    \includegraphics[width=0.35\textwidth]{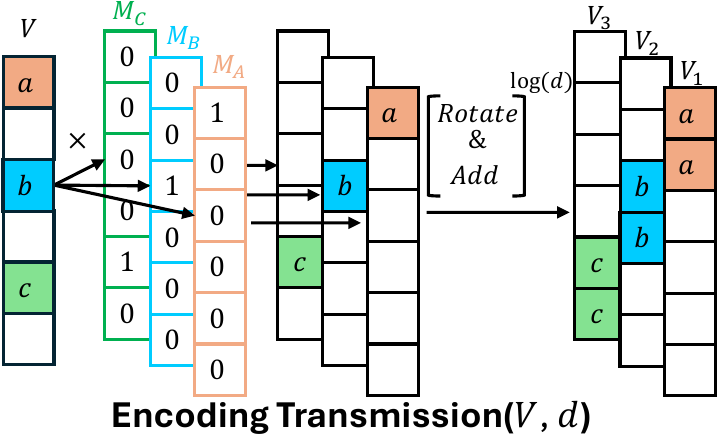} 
    \captionsetup{skip=2pt}
    \caption{Encoding transmission for ARCC inner-inner: masking and rotating an inner-packed vector to broadcast a scalar feature $q_j$ across valid slots, aligning it with an outer-packed cache block before CT$\times$CT multiplication.}
    
    \label{fig:element_duplication}
\end{figure}
Since the $q$ and $a$ utilize inner-based packing, their scalar components are compacted into a single ciphertext. However, the computation requires each valid scalar slot to correspond precisely with an entire inner-based packing structure. To resolve this slot misalignment as depicted in the interaction flow (as shown in Figure \ref{fig:ctct}(b) $Q'\&K$ and Figure \ref{fig:element_duplication}), we must first decouple and replicate the scalars before executing the multiplication steps. Since the ciphertext $q$ contains the full vector, we first isolate each scalar feature $q_j$. This is achieved by applying a plaintext mask to zero out other elements, followed by a specific rotation to replicate $q_j$ across all valid slots. This creates a "broadcasted" ciphertext where every slot holds $q_j$, aligning the scalar with the target matrix structure.
This broadcasted ciphertext is homomorphically multiplied with the corresponding outer-packed $K$. The results are accumulated across the feature dimension $j$ to form the final score vector.

\noindent\textbf{ARCC Inner-Outer Mode.}
This kernel handles the multiplication between an inner-packed operand and an outer-packed operand. In autoregressive decoding, it is used to (i) compute attention scores against the generated history $s_{auto} = q K_{auto}^T$ and (ii) aggregate values for the prefilling history $o_{prefill} = a V_{prefill}$. For $s_{auto}$, the generated cache $K_{auto}$ is physically stored in an inner-packed layout, but attention requires $K_{auto}^T$, which behaves as an outer-packed structure. Therefore, the interaction between the inner-packed $q$ and the outer-packed $K_{auto}^T$ is classified as inner-outer.

To implement this, we leverage the physical alignment between the query and the row-stored keys to perform a parallel Dot-product Reduction as illustrated in Figure \ref{fig:ctct}(c). We first execute a single SIMD multiplication between the inner-packed $q$ and the outer-packed $K$, producing a ciphertext of element-wise products. To aggregate these products into a scalar score without sequential summation, we employ a "folding sum" technique. This involves performing a series of rotations by powers of 2 and accumulating the results. This reduces the summation complexity from linear $O(d)$ to logarithmic $O(\log d)$, fully exploiting the SIMD slots for efficient reduction.

\noindent\textbf{Complexity and Scalability Analysis.}
Table~\ref{tab:complexity_detailed} summarizes the asymptotic complexity of CryptoGen compared to state-of-the-art secure inference baselines adapted for auto-regression. 
\begin{table}[htbp]
\centering
\caption{Complexity of CT$\times$CT attention operations per stage. We report asymptotic rotation and CT$\times$CT costs in prefilling ($m$ tokens) and autoregressive decoding (current length $k$), where $d$ is the hidden dimension and $n$ is the SIMD slot count.}
\label{tab:complexity_detailed}
\resizebox{0.9\columnwidth}{!}{%
\begin{tabular}{|l|c|c|c|c|}\hline

{\textbf{Method}}& \multicolumn{2}{|c|}{\textbf{Prefilling Phase}} & \multicolumn{2}{|c|}{\textbf{Autoregressive Phase}} \\\hline& \textbf{Rotation} & \textbf{CT$\times$CT}& \textbf{Rotation} & \textbf{CT$\times$CT}\\\hline 
BOLT & $O(d)$ & $O(m^2)$ & $O(d)$ & $O(k^2)$\\\hline
THOR & $O(d)$ & $O(m^2)$ & $O(d)$ & $O(k^2)$\\\hline
CryptoGen& $O(d)$ & $O(m^2)$ & $O(\log d)$& $O(k)$\\ \hline
\end{tabular}%
}
\vspace{2pt}
\begin{flushleft}
\footnotesize
\textit{Note:} Baselines (BOLT/THOR) incur $O(L^2)$ complexity in generation due to stateless re-computation of the full attention matrix at each step. CryptoGen reduces this to a linear scan $O(L)$ via stateful caching.
\end{flushleft}
\end{table}
As shown in Table~\ref{tab:complexity_detailed}, BOLT and THOR are statistically bound to their rigid packing layouts (Outer/Diagonal-based). In auto-regression, adding a single token breaks their structure, forcing a "stateless" execution where the entire history must be re-processed or re-encoded at every step $L$. This results in a quadratic $O(L^2)$ cumulative complexity for attention. In contrast, CryptoGen's Heterogeneous KV-Cache and ARCC kernels enable true stateful processing. By computing only the incremental interaction between the new query and the cached history, CryptoGen reduces the per-step workload to a linear scan $O(L)$.

Even for the per-step computation, baselines suffer from redundant computation. BOLT/THOR requires the same linear $O(d)$ rotations with that of prefilling stage. CryptoGen minimizes this overhead using the "Folding Sum" technique within its Inner-Inner/Outer kernels, compressing the accumulation depth to logarithmic $O(\log d)$. This significantly lowers the constant latency factor for both linear projections and attention scoring.

\subsection{Encrypted KV-Cache Management}\label{ss:KVM}
Maintaining a persistent Key-Value (KV) cache in the encrypted domain presents unique challenges that do not exist in plaintext generation.
During auto-regressive generation, the cache must evolve dynamically: historical states from the prefilling phase must be preserved, while new token embeddings are continuously appended.
Two critical bottlenecks arise in this process:
\begin{enumerate}
    \item \textit{Noise Accumulation:} Homomorphic operations (especially multiplications) consume the noise budget of ciphertexts. The newly generated tokens ($k_{m+1}, v_{m+1}$) are themselves the product of deep computations (CT$\times$PT and CT$\times$CT), leaving them with a critically low remaining budget. Simply appending these "noisy" ciphertexts to the cache can quickly push the cumulative noise beyond the decryption threshold, causing correctness failure.
    \item Slot Inefficiency and Memory Explosion: Without optimization, each new token would occupy a fresh ciphertext. Since a single token vector uses only a fraction of the available SIMD slots (e.g., $d=768$ vs $n=8192$), this naive appending leads to severe slot underutilization. This effectively regresses the system to a discriminative-like framework where each generated key-value pair occupies a dedicated ciphertext, mirroring the naive caching limitations discussed in Section~\ref{subsec:motivation_bolt_limit}. Consequently, the number of cache ciphertexts grows linearly with sequence length, triggering a quadratic increase in computational cost for future attention steps and exhausting system memory.
\end{enumerate}

To overcome these barriers, \textit{CryptoGen} introduces a specialized cache management framework that integrates two synergistic strategies (formally described in Algorithm~\ref{alg:cache} in Appendix~\ref{subsec:appendix_alg}):

\textbf{Lazy Noise Refreshing (Addressing Noise).}
To ensure computational stability without incurring excessive communication overhead, CryptoGen adopts a Lazy Refreshing strategy. Instead of refreshing the cache at every step, the server monitors the noise budget and triggers a collaborative refresh protocol only when the budget approaches exhaustion.
As detailed in the appendix, the server masks the noisy ciphertext with a random secret share ($K/V - r$) and sends it to the client. The client decrypts and re-encrypts the data to generate a "fresh" ciphertext, which the server then reconstructs by removing the mask ($K/V_{refresh} + r$). This protocol restores the noise margin without leaking plaintext. By amortizing the refresh cost over many generation steps, the MPC communication overhead becomes negligible.

\textbf{Ciphertext Concatenation (Addressing Inefficiency).}
To maximize memory efficiency, CryptoGen implements a Ciphertext Concatenation protocol that merges multiple token vectors into compact cache blocks. Leveraging the \textit{inner-diagonal encoding}, newly generated keys and values are naturally aligned for packing.
The system calculates a specific slot offset based on the current cache capacity ($B = \lceil n/d_2 \rceil$) and utilizes homomorphic addition to append the masked new token ($k_{m+1}, v_{m+1}$) into the next available slot block of the current cache ciphertext.
This slot-aware accumulation ensures that cache ciphertexts are densely packed. It keeps the total number of ciphertexts nearly constant relative to the sequence length, thereby reducing the complexity of the $QK^T$ attention operation from what would be a linear growth under naive caching to a significantly flatter curve.

By combining these mechanisms, CryptoGen reduces both the ciphertext count and update latency by more than one order of magnitude compared to baseline approaches, enabling stable and scalable long-context generation.

%% file: contents_usenix/Experimental_Methodologies.tex
\subsection{Experimental Setup}
All experiments are conducted on a workstation equipped with an AMD Ryzen Threadripper PRO 3955WX CPU featuring 16 physical cores (32 threads), running at a base frequency of 2.20~GHz with up to 3.90~GHz boost frequency. The machine is provisioned with 125~GiB of system memory and runs Ubuntu~22.04.5~LTS (Linux kernel~6.8.0-85-generic). CryptoGen follows the standard two-party semi-honest MPC setting as in prior work. %To isolate cryptographic overhead from physical network variability, our experiments utilize LAN environment for communication between the client and server processes. \ql{why only LAN, why not WAN?} 
The code is compiled using \texttt{gcc}~11.4.0 and \texttt{g++}~11.4.0 . Plaintext baselines are executed on the same machine under identical software configurations to ensure a fair comparison.

\subsection{Cryptographic Libraries and Parameters}

CryptoGen is implemented as a hybrid HE--MPC framework. For homomorphic encryption, we use the Microsoft SEAL library\cite{seal2020,laine2017seal} and adopt the BFV scheme\cite{brakerski2012fully,fan2012somewhat}, which supports exact integer arithmetic and is well suited for our fixed-point quantized Transformer model. The HE parameters follow SEAL’s recommended settings for 128-bit security. We use a polynomial modulus degree of $n = 8192$.

The model weights and activations are converted into fixed-point integers. We use a plaintext modulus of approximately $p \approx 2^{29}$, which accommodates the dynamic range of GPT-2 activations after quantization. The ciphertext modulus is chosen to be $q \approx 2^{220}$, providing a sufficiently large noise budget to support the continuous multiplication of ciphertext-plaintext and ciphertext-ciphertext in auto-regressive generation. And the larger ciphertext modulus ensures stable encrypted computation without frequent noise refreshing with using encrypted KV cache.

For nonlinear components such as GELU, LayerNorm, and softmax, CryptoGen employs the EzPC \cite{chandran2019ezpc,rathee2022secfloat,rathee2020cryptflow2}secure computation library, running a standard two-party semi-honest MPC protocol. All nonlinearities are executed in the MPC domain, while linear layers—including attention projections and feed-forward transformations—are performed using HE operations. This division of labor mirrors prior hybrid approaches while enabling efficient and noise-stable encrypted decoding.

\subsection{Model and Datasets}

We evaluate CryptoGen using the GPT-2 base architecture, which consists of 12 Transformer decoder layers with a hidden size of 768, 12 attention heads, and GELU activations. \\
\paragraph{Accuracy \& Efficiency.}
We evaluate the perplexity of the Wikitext-2\cite{merity2016pointer}, Penn Treebank (PTB)\cite{marcus1993ptb}, and LAMBADA\cite{paperno2016lambada} test datasets with and without using CryptoGen. We also compare the runtime of CryptoGen and BOLT when generating sequences of different lengths.

Since BOLT is originally designed for discriminative Transformer inference, we adapt it to the autoregressive setting. Specifically, we pre-allocate a sequence of placeholder tokens equal to the desired generation length. %\ql{BOLT-C looks confusing for reviewers, we may not explicitly use  the -c}
During generation, each newly predicted token replaces one placeholder, and the prefix combined with all previously generated tokens is fed into the model to predict the next token. This approach enables BOLT to perform auto-regressive decoding while preserving its original HE-MPC execution pipeline and packing strategy. Both CryptoGen and BOLT use a fixed prefix of 64 tokens taken from the corresponding test sets. We measure decoding latency across generation lengths ranging from 64 to 512 tokens, generating one token at a time in an autoregressive manner.

\subsection{Non-linear Layer Computation}
For the nonlinear operations required in Transformers, such as GELU activations and Softmax normalization, \textit{CryptoGen} employs MPC protocols. This hybrid approach avoids the costly polynomial approximations typically required in FHE-only systems. We adopt the efficient MPC protocols of BOLT \cite{pang2024bolt}, which are optimized for the semi-honest two-party setting. The overview of performing nonlinear computations within this hybrid framework is as follows: To securely switch from HE to MPC, the server first generates a random plaintext mask and homomorphically subtracts it from the input HE ciphertext. This blinded ciphertext is then sent to the client for decryption. This process effectively splits the input into two secret shares: the client holds the decrypted blinded value, while the server holds the random mask. Privacy is preserved as neither party can reconstruct the true input value. Subsequently, the client and the server jointly execute the designated MPC protocol on their respective shares. To convert the result back to an HE ciphertext, the client encrypts the resulting output share and sends it to the server. The server then homomorphically adds its own output share (derived from the MPC computation on its initial mask) to the client's ciphertext, yielding the final encrypted result of the nonlinear function. 

\noindent\textbf{GELU and Polynomial Optimization.} For activations like GELU, \textit{CryptoGen} utilizes the highly optimized 4-degree polynomial approximations. These approximations exploit properties of the GELU function to reduce the computational domain. First, GELU behaves linearly ($GeLU)x\approx x$ or $GeLU)x\approx 0$) outside a small input range. The polynomial approximation is therefore only applied within this critical range ([-3.2, 3.2])\cite{kim2021bert}. Second, the protocol leverages the symmetry of $GeLU$ function, allowing the polynomial to be evaluated only on positive inputs, further halving the approximation domain. To minimize the communication cost of evaluating this 4-degree polynomial, the 4-degree polynomial can be reformulated as a composition of two 2-degree polynomials. By doing this, a sequential multiplication step can be skipped, saving one round of MPC protocol communication. 

\noindent\textbf{Softmax.} For the computationally expensive Softmax function, \textit{CryptoGen} implements the efficient integer-only protocol, which avoids expensive Look-Up Tables. After normalizing the inputs by subtracting $x_{max}$, the core mechanism securely decomposes the input $\tilde{x_i}$ into $\tilde{x_i} = (-ln~2) \cdot z + p$. This decomposition is achieved by first computing the integer $z$ using a secure fixed-point multiplication and a right-shift (to simulate division by $-ln2$). The value $p$ is then computed from $z$ and $\tilde{x_i}$. Because $p$ is guaranteed to be in a small fixed range ($(-ln2, 0]$) , the $exp(p)$ term can be accurately and cheaply computed with a simple 2-degree polynomial. The final $exp(\tilde{x_i}) = exp(p) >> z$ result is calculated by applying the polynomial's output to a sequence of secure multiplexers controlled by the secret-shared bits of $z$. The final division in Softmax is handled using an established secure reciprocal protocol from SIRNN\cite{rathee2021sirnn}.

%% file: contents_usenix/6_result.tex
\subsection{End-to-End Performance}\label{subsec:end2end}
Figure~\ref{fig:scalability} compares the end-to-end runtime of CryptoGen and BOLT under autoregressive inference. Due to BOLT's packing scheme, the effective sequence length must align with the ciphertext block size. When the number of generated tokens does not fill an entire block, BOLT must pad the input, producing the staircase-shaped latency curve observed in the figure. Each step increase results in a sharp latency jump, and the overall computation scales quadratically with the sequence length.

In contrast, CryptoGen exhibits a linear growth pattern. Its ciphertext-ciphertext multiplications in the $QK^T$ computation involve only a single new ciphertext $Q$ at each auto-regressive step, making the scaling behavior highly predictable. More importantly, CryptoGen maintains an encrypted KV cache throughout generation, eliminating the need to recompute all past key and value vectors at every step. This reduces the attention cost from $O(L^2)$ to $O(L)$ throughout the entire decoding process. In addition, unlike BOLT—which must reprocess the full prefix and all generated tokens at every step—CryptoGen only encrypts and processes one new token per step. As a result, the number of ciphertexts involved in the linear layers and attention modules remains nearly constant, also reducing both computation and communication in MPC-heavy components such as \textit{softmax} and GeLU. Consequently, CryptoGen demonstrates substantially better end-to-end scaling and improves per-step latency by 4.4$\times$ to 7.6$\times$ over BOLT as the sequence length increases.

In addition to runtime, Figure~\ref{fig:motivation} also reports the memory usage during autoregressive decoding. The memory behavior of the two systems differs substantially. BOLT exhibits a staircase-shaped memory pattern that mirrors its runtime curve. Because BOLT must repeatedly pad the input to match the ciphertext block size, each increase in sequence length expands the number of packed ciphertexts it maintains. This causes both the intermediate ciphertext buffers and the attention-related ciphertext matrices to grow proportionally, leading to sharp memory spikes at each block boundary. Furthermore, BOLT recomputes all key and value vectors at every decoding step, producing a large volume of transient ciphertexts whose size increases with the prefix length.

In contrast, CryptoGen maintains a significantly more stable memory footprint across all generation lengths. Its encrypted KV cache avoids regenerating past key and value ciphertexts, preventing quadratic growth in intermediate tensors. CryptoGen’s inner-encoding scheme and encrypted KV cache concatenation also keep the number of ciphertexts per layer nearly constant, independent of the current prefix length. As a consequence, the memory cost of attention and feed-forward layers remains flat as the sequence length increases. Additionally, because CryptoGen processes only one new token per iteration, it avoids the need to store full-sequence ciphertext batches, resulting in substantially lower and smoother memory consumption. In general, CryptoGen achieves both lower peak memory and more predictable usage patterns than BOLT, further highlighting the scalability benefits of its hybrid HE–MPC design.

\begin{table}[htbp]
\centering
\resizebox{1\columnwidth}{!}{%
\begin{tabular}{|c|c|c|c|}\hline
 & \textbf{Wikitext2} & \textbf{PTB} & \textbf{LAMBADA} \\\hline
Plaintext & $30.63\pm 0.031$ & $53.65\pm 0.045$ & $55.86\pm 0.091$ \\\hline
\textbf{CryptoGen} & 30.66 & 54.64 & 55.92 \\\hline
\end{tabular}
}\captionsetup{skip=2pt}
\caption{Perplexity of Plaintext and CryptoGen inference.}
\label{tab:ppl_res}
\end{table}

Table~\ref{tab:ppl_res} further reports the perplexity of CryptoGen compared with plaintext GPT-2 inference on Wikitext-2, PTB, and LAMBADA. Across all three datasets, CryptoGen preserves model accuracy with negligible degradation. The perplexity values differ from plaintext by less than a small margin, demonstrating that our fixed-point integer representation and hybrid HE-MPC computation do not introduce meaningful numerical drift. These results confirm that CryptoGen maintains the same generation quality as the original model while enabling fully privacy-preserving inference.

\subsection{Computation Algorithm Comparison}
\begin{figure}[htbp]
    \centering
    \includegraphics[width=0.4\textwidth]{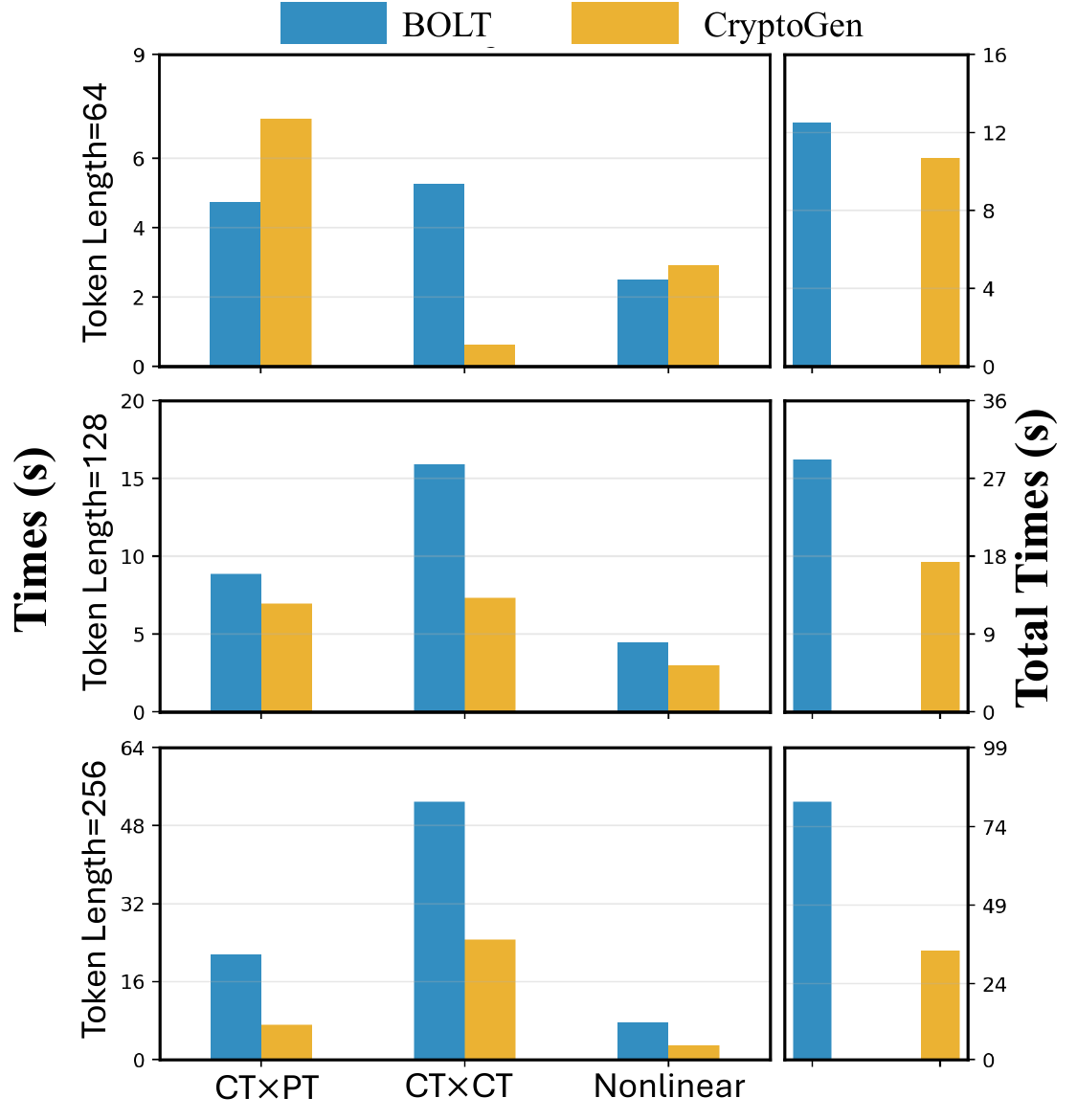} 
    \captionsetup{skip=2pt}
    \caption{Breakdown end-to-end inference between CryptoGen \& BOLT}
    \label{breakdown_compare}
\end{figure}
Figure~\ref{breakdown_compare} presents a detailed breakdown of the runtime of CryptoGen and BOLT at token length 64,128,256. We categorize the execution time into three primary components: ciphertext-plaintext multiplications (CT×PT), ciphertext-ciphertext multiplications (CT×CT), and MPC-based nonlinear operations. The two systems demonstrate different scaling behaviors as the sequence length grows.

From the perspective of component proportions, we observe that when the token length is small (e.g. $L = 64$), the runtime contributions of CT×PT and CT×CT in BOLT are relatively close. However, as the sequence length increases, the CT×CT component grows significantly faster than CT×PT. This behavior is expected: CT×CT operations correspond to $QK^T$ and $\mathrm{softmax} \cdot V$ computations, whose computational cost scales quadratically with the sequence length in non–autoregressive HE execution. In contrast, CT×PT operations originate from the linear layers, whose complexity grows linearly with $L$, resulting in much slower growth.

CryptoGen exhibits a different pattern. Although the proportion between CT×PT and CT×CT also changes with increasing token length, the CT×PT cost remains nearly constant across all sequence lengths. This is because CryptoGen processes only a single new ciphertext token at each auto-regressive step, so the CT×PT workload does not depend on the current prefix length. The only component that grows with $L$ is the CT×CT portion, which reflects the inherent length-dependent attention cost. Consequently, the variation in CryptoGen’s breakdown is primarily due to the linear growth of CT×CT, while CT×PT remains stable regardless the length of sequence .

\paragraph{Nonlinear MPC operations.}
A similar trend can be observed in the MPC-based nonlinear components. BOLT repeatedly processes the full prefix and generated tokens, causing the number of secret shares and communication rounds to increase with $L$. CryptoGen processes only one new token per iteration, which significantly reduces the volume of MPC communication. Consequently, CryptoGen exhibits nearly flat nonlinear cost curves, while BOLT’s nonlinear overhead grows steadily as the sequence length increases.

\subsection{Breakdown Analysis of a Single Transformer Block}
\begin{table}[htbp]
\centering
\scriptsize
\setlength{\tabcolsep}{3pt}
\resizebox{\columnwidth}{!}{%
\begin{tabular}{|c|c|c|c|c|c|c|c|c|}
\hline
\textbf{Token Length} & \multicolumn{2}{|c|}{\textbf{64}} & \multicolumn{2}{|c|}{\textbf{128}} & \multicolumn{2}{|c|}{\textbf{256}} & \multicolumn{2}{|c|}{\textbf{512}} \\\hline
\textbf{Method}& \textbf{BOLT}& \textbf{CryptoGen}& \textbf{BOLT}& \textbf{CryptoGen}& \textbf{BOLT}& \textbf{CryptoGen}& \textbf{BOLT}& \textbf{CryptoGen}\\\hline
$x*W$ (CT$\times$PT) / s& 1.77& 0.74& 3.21& 0.86& 9.90  & 0.74    & 26.54  & 0.64   \\\hline
Linear 2 (CT$\times$PT) / s& 0.37& 0.60  & 0.64 & 0.57 & 1.12 & 0.55   & 2.35  & 0.52  \\\hline
Linear 3 (CT$\times$PT) / s& 1.35  & 2.80  & 2.50  & 2.85  & 4.63  & 2.97  & 7.35  & 2.64  \\\hline
Linear 4 (CT$\times$PT) / s& 1.26  & 2.76  & 2.51  & 2.68  & 5.86  & 2.90  & 8.28  & 2.76  \\\hline
$QK^{\top}$(CT$\times$CT) / s& 2.47    & 0.27     & 7.09    & 3.55  & 24.86  & 12.25   & 77.23  & 28.27  \\\hline
Softmax * V (CT$\times$CT) / s& 2.80  & 0.35 & 8.80  & 3.76  & 28.09 & 12.32   & 95.47  & 28.41   \\\hline
Softmax (Nonlinear) / s& 0.57 & 0.17   & 1.57  & 0.19  & 3.77  & 0.24  & 4.97  & 0.29  \\\hline
GeLU / s& 0.75 & 1.12   & 1.27  & 1.12   & 1.80  & 1.07  & 2.23  & 1.18   \\\hline
Norm / s& 0.58 & 0.79   & 0.83 & 0.87  & 1.02 & 0.80  & 1.28  & 0.84  \\\hline
 Total / s& 11.92& 9.6& 28.42& 16.45& 81.05& 33.84& 225.7&65.55\\\hline

\end{tabular}
}
\captionsetup{skip=2pt}
\caption{Runtime Breakdown of a Single Transformer Block Across Different Sequence Lengths for BOLT and CryptoGen }
\label{tab:breakdown}
\end{table}
Table~\ref{tab:breakdown} presents the detailed runtime breakdown of a single Transformer block under different token lengths ($L=64, 128, 256, 512$) for both BOLT and CryptoGen. Each row corresponds to a major computational component, including CT$\times$PT,CT$\times$CT, and MPC-based nonlinear operations. The breakdown reveals how different subsystems scale with sequence length and highlights the root causes of the performance divergence between the two frameworks.

\noindent \textbf{CT$\times$PT Components (Linear 2/3/4).}
Across all sequence lengths, CryptoGen maintains nearly constant CT$\times$PT runtime. This is because CryptoGen processes only \emph{one} new ciphertext token at each auto-regressive step; thus, the workload of CT$\times$PT linear layers is independent of the current prefix length. As a result, CryptoGen’s CT$\times$PT cost remains flat even as the sequence length grows from 64 to 512.

For BOLT, CT$\times$PT runtime increases steadily with sequence length. Since BOLT processes the entire token length at each generation step, the number of input ciphertexts grows linearly with $L$. Consequently, the CT$\times$PT time of BOLT roughly triples from $L=64$ to $L=256$ and increases by more than 15$\times$ at $L=512$.

\noindent \textbf{CT$\times$CT Components ($QK^\top$ and Softmax$\cdot V$).}
The most significant discrepancy appears in CT$\times$CT operations. As the sequence length increases, BOLT exhibits extremely rapid growth in both the $QK^\top$ and $\mathrm{softmax}\cdot V$ stages. For instance, the $QK^\top$ time rises from 2.47\,s at $L=64$ to 77.23\,s at $L=512$ (a 31$\times$ increase), while $\mathrm{softmax}\cdot V$ grows from 2.80\,s to 95.47\,s (a 34$\times$ increase). This behavior comes from BOLT’s non-autoregressive execution model: without encrypted KV caching, both $Q$ and all historical $K$/$V$ vectors must be recomputed at each step, resulting in a \emph{quadratic} time complexity in the sequence length ($O(L^2)$). At long sequences, CT$\times$CT operations dominate BOLT’s total runtime.

CryptoGen, on the other hand, exhibits only \emph{linear} growth in CT$\times$CT runtime. With encrypted KV caching, CryptoGen performs CT$\times$CT operations only between the new $Q$ ciphertext and cached K/V vectors. Thus, the CT$\times$CT cost grows approximately proportionally with $L$ (e.g., $\mathrm{softmax}\cdot V$: 0.35\,s $\rightarrow$ 28.41\,s), avoiding the quadratic explosion seen in BOLT.

\noindent \textbf{MPC Nonlinear Components.}
The MPC-based nonlinear operations also show distinct trends. In CryptoGen, the nonlinear workload remains mostly constant since these operations only involve the new token. In BOLT, however, the nonlinear runtime increases moderately with sequence length due to the expanding ciphertext representations that must participate in MPC. Although the growth is not as dramatic as CT$\times$CT, it contributes noticeably to BOLT’s total runtime at larger $L$.

Furthermore, it is worth noting the impact of our noise management strategy. Due to the Lazy Refreshing policy, the communication overhead of the noise refresh protocol is effectively amortized over the entire generation process. In our evaluation with sequence lengths up to 512, the overhead remains negligible, contributing less than 1\%.

\paragraph{Total Runtime per Layer.}
Summing all components highlights the cumulative effect of the above trends. As shown in Table~\ref{tab:breakdown}, BOLT’s total per-layer latency increases from 11.92\,s at $L=64$ to 225.7\,s at $L=512$, while CryptoGen increases from 9.6\,s to 65.55\,s. CryptoGen therefore achieves a 3.4$\times$ speedup at $L=512$, and maintains far more predictable scaling due to linear-time attention and constant-time linear layers. In contrast, BOLT’s quadratic attention cost quickly dominates, becoming the primary bottleneck at longer sequence lengths.

% \begin{figure}[htbp]
%     \centering
%     \includegraphics[width=0.5\textwidth]{figure/runtime_64_512.pdf} 
%     \captionsetup{skip=2pt}
%     \caption{End-to-end runtime and input ciphertext number for each step between CryptoGen LLM \& BOLT}
%     \label{res_compare_bolt}
% \end{figure}% 右轴

%% file: contents_usenix/7_conclusion.tex
In this paper, we presented CryptoGen, a hybrid HE-MPC framework that enables efficient and privacy-preserving auto-regressive Transformer generation. By combining heterogeneous encoding for prefilling and generation stage and two complementary ciphertext-ciphertext multiplication, CryptoGen fundamentally changes the scaling behavior of secure generation. Our design eliminates the quadratic overhead inherent in prior HE-only approaches such as BOLT and reduces the complexity of attention from $O(L^2)$ to $O(L)$ in auto-regressive process.

We implemented CryptoGen on GPT-2 and conducted extensive evaluations on Wikitext-2, PTB, and LAMBADA. Experimental results show that CryptoGen preserves model accuracy, achieving perplexity nearly identical to plaintext inference. In end-to-end runtime, CryptoGen demonstrates substantially better scalability than BOLT: while BOLT suffers from quadratic growth due to repeated recomputation of all prefix tokens, CryptoGen maintains linear-time latency and achieves $4.4\times$--$7.6\times$ speedup at longer sequence lengths. Our component-level breakdown further confirms that CryptoGen keeps CT$\times$PT and nonlinear MPC operations constant per step, while attention-related CT$\times$CT operations grow only linearly.

Overall, CryptoGen provides a new path toward practical, privacy-preserving autoregressive generation by heterogeneously combining different stages of the decoding pipeline, effectively reducing ciphertext volume, minimizing MPC communication, and ensuring predictable linear scaling. In future work, we plan to extend CryptoGen to larger language models, explore more advanced ciphertext packing strategies, and investigate integration with GPU-accelerated secure computation to further improve performance.

%% file: contents_usenix/appendix.tex
\subsection{KV Cache Management}\label{subsec:appendix_alg}
\begin{algorithm}[H]
\small
\caption{Noise Refresh and Concatenation of $K/V^{auto}$}
\label{alg:cache}
\begin{algorithmic}[1]
\REQUIRE Current caches $K^{auto}_m, V^{auto}_m$; \\new entries $k^{auto}_{m+1}, v^{auto}_{m+1} \in \mathbb{Z}_p^{1\times d_2}$;\\ slot count $n$
\ENSURE Updated caches $K^{auto}_{m+1}, V^{auto}_{m+1}$

\STATE \textbf{Step 1: Noise Budget Check}
\IF{Cache noise budget exhausted}
    \STATE Server samples a random plaintext vector $r$
    \STATE Server computes masked share:
    \[
        K/V^{auto}_{share} = K/V^{auto}_m - r
    \]
    \STATE Server $\rightarrow$ Client: send $K/V^{auto}_{share}$
    \STATE Client decrypts and re-encrypts:
    \[
        K/V^{auto}_{refresh} = K/V^{auto}_{share}
    \]
    \STATE Client $\rightarrow$ Server: send $K/V^{auto}_{refresh}$
    \STATE Server reconstructs refreshed ciphertext:
    \[
        K'/V'^{auto}_m = K/V^{auto}_{refresh} + r
    \]
    \STATE Replace original caches with refreshed versions:
    \[
        K/V^{auto}_m = K'/V'^{auto}_m
    \]
\ENDIF
\STATE $B = \lceil n/d_2 \rceil$
\STATE $pos(m{+}1) = ((m{+}1) \bmod B) \cdot d_2$
\STATE \textbf{Step 2: Cache Concatenation (Key)}

\STATE $ct\_new = \mathrm{MultPlain}(k^{auto}_{m+1}, mask_{pos(m+1)})$ 
\STATE $K^{auto}_{m+1} = K^{auto}_m+ ct\_new$

\STATE \textbf{Step 3: Cache Concatenation (Value)}
\STATE $ct\_new \gets \mathrm{MultPlain}(v^{auto}_{m+1}, mask_{pos(m+1)})$
\STATE $V^{auto}_{m+1}= V^{auto}_m+ ct\_new$

\RETURN $K^{auto}_{m+1}, V^{auto}_{m+1}$
\end{algorithmic}
\end{algorithm}

To ensure the scalability and correctness of long-sequence generation, CryptoGen implements a dual-strategy management framework as formalized in Algorithm 1. This algorithm addresses two primary bottlenecks: homomorphic noise explosion and memory inefficiency.\\
1. \textit{Lazy Noise Refreshing} (Lines 1-10) \\
The "Lazy" strategy minimizes communication overhead by triggering a collaborative refresh only when the ciphertext's noise budget approaches the decryption threshold.
\begin{itemize}
    \item The server applies a random mask $r$ to the noisy cache $K/V_{m}^{auto}$ and sends the blinded share to the client.
    \item The client decrypts and re-encrypts the data, effectively resetting the noise budget without learning the underlying plaintext. This amortization ensures that the refresh cost becomes negligible over long generation steps. 
\end{itemize}
2. \textit{Slot-Aware Ciphertext Concatenation} (Lines 11-19) \\
To prevent memory explosion, we avoid storing each new token in a dedicated ciphertext.  
\begin{itemize}
    \item Leveraging inner-diagonal encoding, the system calculates the optimal slot offset $pos(m+1)$ based on the current cache capacity $B = \lceil n/d_2 \rceil$.
    \item New keys $k_{m+1}$ and values $v_{m+1}$ are homomorphically added into the next available slots of the existing cache ciphertext. This mechanism keeps the total number of ciphertexts nearly constant, reducing the attention complexity from linear growth to a significantly flatter curve.
\end{itemize}